\documentclass[fleqn,twoside]{article}
\usepackage{espcrc1}

\newcommand{\AmS}{{\protect\the\textfont2
  A\kern-.1667em\lower.5ex\hbox{M}\kern-.125emS}}
\title{Towards combinatorial modeling of  wireless technology
  generations}

\author{Mark Sh. Levin
\address{Inst. for Information Transmission Problems,
 Russian Academy of Sciences\\
 19 Bolshoj Karetny Lane, Moscow 127051, Russia\\
 }
\thanks{
  {\it E-mail address}: mslevin@acm.org
  }
 }

\begin{document}

\maketitle

\begin{abstract}
 The paper addresses the following problems:
 (1) a brief survey on wireless mobile communication technologies
 including evolution, history evolution
  (e.g., chain of system generations 0G, 1G, 2G, 3G, 4G, 5G, 6G, 7G);
 (2) using a hierarchical structural modular approach
  to the generations of the wireless communication systems
  (i.e., hierarchical combinatorial modeling of the communication technologies),
 (3) illustrative usage of two-stage combinatorial approach
  to improvement/forecasting of the communication technology
 (a version of 5G) (on the basis of multiple choice problem).
 Numerical examples illustrate the suggested combinatorial approach.

~~~


\vspace{1pc}
\end{abstract}

\newcounter{cms}
\setlength{\unitlength}{1mm}

\tableofcontents

\newpage
\section{Introduction}

 Recently, in many engineering domains
 several technology generations
 are under study, design and forecasting  processes.
 Some examples of applied system generations are pointed out in Table 1.
 As a result,
 special methods have to be suggested and used for the modeling
 the system generations changes/evolution and system forecasting.

\begin{center}
 {\bf Table 1.} Examples of technology generations \\
\begin{tabular}{| c | l| l |l |}
\hline

 No.& Domain &Generations & Some source(s) \\
\hline
 1.&Fighter aircraft
   &\(1\), \(2\),  \(3\), \(4\), \(4^{+}\), \(4^{++}\), \(5\), \(6\)
  &\cite{chie14,heb08,tir09}\\

 2.&Partiot air and missile&
 Patriot, Patriot PAC-2, Patriot PAC-3&\cite{andrew10,hild07,sandru13,sher03}\\

 &defense systems (USA)&&\\

 3.&Anti-aircraft weapon systems
 &S75,  S125, S200,  S300, S400, S500&\cite{andrew10,bart95,efr94} \\

  &(Russia)&&\\

 4.&Anti-aircraft weapon systems
 &``Krug'',  S300V, S300VM, S300V4&\cite{andrew10,bart95,efr94} \\

  &for ground forces (Russia)&&\\

 5.&Standard for  multimedia
    & MPEG, MPEG-2, MPEG 4& \cite{lev15,levinf09}\\

   & information transmission&&\\

 6.&Wireless mobile
    & 0G, 1G, 2G, 3G, 4G, 5G, 6G, 7G
 &\cite{berez02,fag14,gill15,gos15,hoss13}\\

 &communication& &\cite{mehta14,sharma13,singh07,yadav17}\\

\hline
\end{tabular}
\end{center}

 In recent years, many research studies are targeted to
  issues of history and evolution of networking, challenges and forecasting
  in communications
  (including  network architecture, network functions, network topology, etc.)
  (Table 2).

\begin{center}
 {\bf Table 2.} Studies of wireless communication systems:
 evolution, history evolution, forecasting \\
\begin{tabular}{| c |  l |l |}
\hline

 No.& Study & Source(s) \\
\hline
 1.&Modeling of topology evolutions and implication on proactive
   routing
     &\cite{wu08}\\
   &overhead in MANETs &\\

 2.&Survey on generations of mobile wireless technology &\cite{bhala10a}\\

 3.&Evolution: 2G, 3G, 4G; challenges in the migration to 4G mobile
 systems
    &\cite{hui03}\\

 4.&Technological changes in the innovation system
    towards 4G mobile service
    &  \cite{sur10} \\

 &(including evolution: 1G, 2G, 3G, 4G)&\\

 5.&Evolution to 4G cellular systems (LTE-Advanced)&\cite{aky10}\\

 6.&Comparative study of 3G and 4G in mobile technology&\cite{kumara11}\\

 7.&Evolution of wireless communication (1G, 2G, 3G, 4G, 5G;
    &\cite{abr16}\\

   &  shift toward green  communication)&\\

 8.&Evolution of mobile wireless communication networks
  (1G, 2G, 3G, 4G, 5G)&\cite{chengx14,patel12a,sharma13}\\

 9.&Comprehensive study about evolution of different mobile generation
   &\cite{mehta14} \\

   &technologies (0G to 5G mobile technology:  survey)&\\

 10.&Comparative studies on 3G, 4G and 5G wireless technology
    &\cite{fag14}\\

 11.&5G on the horizon: key challenges for the radio-access network
   &\cite{deme13}\\

 12.&Radio access network evolution&\cite{vae17}\\

 13.&Study of next generation wireless network 6G&\cite{singh07}\\

 14.&Upcomming technologies: 5G and 6G&\cite{gos15}\\

 15.&Future of wireless technology 6G \& 7G &\cite{khu15}\\

 16.&Challenges and opportunities
   for next generation of mobile networks
      & \cite{chin14,hanb15,haw14}\\

 17.&Challenges and evolution of next generations wireless communication
   &\cite{yadav17}\\

 18.&Combinatorial approach for modular system evolution:&\cite{lev02,lev06,lev13,lev15}\\

 18.1.&Combinatorial evolution of MPEG-like standard for multimedia information

     &\cite{lev15,levinf09}\\

 18.2.&Combinatorial evolution and forecasting of communication protocol ZigBee
      & \cite{lev15,levsib10,levand12}\\

  18.3.&Combinatorial evolution and forecasting of requirements
   to  network topology&\cite{lev17evo}\\

 19.& Historical evolution of software defined networking (SDN), its architecture
    &\cite{kreutz15,singh16}\\

 20.&Intellectual history of programmable networks (SDN)
  & \cite{feam14,nunes14}\\

 21.&History and challenges in network function virtualization
   &\cite{chow10,mij16,vae17}\\

\hline
\end{tabular}
\end{center}

 In general (from the viewpoint of innovation processes),
 the following four classes of the impacts of new technologies in communications
 can be considered
   \cite{afuah03,bocc14,lev10sib,lev13home,lev15}:

 {\bf 1. Local evolutions/modifications}  (minor changes,
 e.g., new coding, signaling support for a higher number of
 antennas):
 {\it 1.1.} at the node level,
 {\it 1.2.} at the architectural level.

 {\bf 2. Component changes:}
  disruptive changes in the design of a class of network nodes
 (e.g., new waveform).

 {\bf 3. Architectural changes}
 (disruptive changes in the system architecture):
 {\it 3.1.} changes of nodes/node functions
 (e.g., new types of nodes or new functions in existing ones),
 {\it 3.2.} changes of system architecture
 (e.g., system topology),
 {\it 3.3.} system extension (e.g., addition of a new system part).

 {\bf 4. Radical changes:}
  disruptive changes that have an impact at both the node
  and the architecture levels.

 Tables 3 and Table 4 contain
  improvement technological directions in wireless mobile systems.

\begin{center}
 {\bf Table 3.} Improvement technological directions in wireless mobile systems, part I \\
\begin{tabular}{| c |  l | l |}
\hline

 No.& Technological direction & Source(s) \\

\hline
 I.& Development directions in 5G (movement from 4G to 5G):
    &\cite{mae16,tikh13}   \\

 1.1.&New radio-interfaces, methods for sending and receiving:&\\

 1.1.1.&new methods of frequency modulation&\\
 1.1.2.&coordination of pomeh based on prospective receiver methods&\\

 1.2.&New network architecture:&\\

 1.2.1.&building of small sots (one point - one user)&\\

 1.2.2.& central architecture:&\\
   &(a) cloud radio-access networks (RAN) based on SDR&\\

   &  and coordinated central controllers,&\\

   & (b) cloud basic networks CN based on SDN&\\

 1.2.3.&multidimensional antennas MIMO&\\

 1.2.4.&evolution technologies MIMO: MIMO active/3D antennas&\\

 1.2.5.& physical division of traffic and control between&\\
    &layers of control and information transmission&\\

 1.3.&New principles and conditions of spectrum usage:&\\

 1.3.1.&usage of new frequency bandwidth (since 6 to 60 GHz)&\\

 1.3.2.&flexible common usage of frequency resources&\\

 1.4.& Smart and adaptive communication networks:&\\

 1.4.1.&usage of mobile applications with their optimization for decreasing &\\

  &the cost of radio-access&\\

 1.4.2.&allocation and management of resources in heterogeneous networks&\\

 1.4.3.&inter-network joint work for different radio-access
 technologies&\\

 1.4.4.&self-adaptation and self-optimization networks&\\

\hline
 II.&Disruptive technology directions for 5G:& \cite{bocc14}\\

 2.1.& Device-centric architectures&\\

 2.2.& Millimeter wave (mmWave)&\\

  2.3.& Massive-MIMO&\\

  2.4.& Smarter devices&\\

  2.5.&Native support for Machine-to-Machine (M2M) communication&\\

\hline
 III.&5G technology advantages: &\cite{kadir15}\\

 3.1.& {\it Category 1} (real time performance): fast response time, low jitter,&\\
  &  latency and delay, and high availability.&\\

 3.2.& {\it Category 2}  (critical infrastructure):
          high reliability, priority access, &\\
   & very wide area coverage.&\\

 3.3.& {\it Category 3} (very high speed broadband):
      Gigabit data rates,  &\\
   &high quality coverage,  multispectrum service.&\\

 3.4.&{\it Category 4} (IoT, M2M):
     more connected devices, deep indoor coverage,&\\
  & signaling efficiency.&\\

 3.5.& {\it Category 5} (virtualized infrastructure): Software defined network,&\\
    & scalable, low-cost systems.&\\

\hline
\end{tabular}
\end{center}

\newpage
\begin{center}
 {\bf Table 4.} Improvement technological directions in wireless mobile systems, part II \\
\begin{tabular}{| c |  l | l |}
\hline

 No.& Technological direction & Source(s) \\

\hline
 IV.&Advantages of 5G as future preffered network:&\cite{fag14}\\


 4.1.&User personalization (high data transfer rates, access
    to large repository &\\

  &of data and services, flexibility)&\\

 4.2.&Terminal and network heterogeneity
    (different types of access networks, &\\
    & e.g., WiMAX, WiFi, UMTS)&\\

 4.3.&High performance (wireless download speeds) &\\

 4.4.& Interoperability (unified global standard, global mobility
   and service &\\

  &portability,
  i.e., different services from different service providers)&\\

 4.5.&Intelligent networking
  (hybrid networks utilizing both the Wireless LAN  &\\

 &concept and WAN design,
  ubiquitous network coverage to users at high speed)&\\

 4.6.&Network convergence
   (convergence with both devices and services) &\\

 4.7.& Lower power consumption&\\

\hline
 V.&Seven technical directions/advantages aiming at augmentation  &\cite{deme13}\\
  &of the wireless world's intelligence:&\\
 5.1.& RAT evolution, &\\
 5.2.& cell shrinking,&\\
 5.3.& composite wireless infrastructure,&\\
 5.4.& heterogeneous networks,&\\
 5.5.& flexible spectrum management,&\\
 5.6.& exploiting cloud concept,&\\
 5.7.& introduction to intelligence&\\

\hline
 VI.&Basic advantages of 6G technology:
    &\cite{gos15,haw14,khu15} \\

 6.1.&Ultra fast access of Internet&\cite{singh07,yadav17}\\
 6.2.&Data rate - up to 10-11 Gbps&\\
 6.3.&Home automation and related applications&\\
 6.4.& Smart homes, smart cities, smart villages&\\
 6.5.&Home based ATM systems&\\
 6.6.&Satellite to satellite communication&\\

\hline
\end{tabular}
\end{center}

 Traditionally,
 studies of  change/improvement/evolution for applied technical systems
 are based on engineering analysis of existing technologies and
 basic trends for future systems improvements.
 Recently, an approach of
 combinatorial modeling for  evolution and forecasting
 of modular systems has been suggested by the author
 \cite{lev98,lev02,lev06,levprob07,lev13,lev15,lev17evo}.
 This combinatorial approach is based on the following:
 (i) hierarchical modular presentation
  (as tree-based morphological structure with alternatives for leaf vertices)
  of the examined system,
 (ii) analysis of system changes or change items/operations
 (e.g., between neighbor system
 generations),
 (iii) assessment of the system changes (change items/operations)
 upon criteria (e.g., cost, prospective utilities);
 (iv) modeling of the system improvement/forecasting process as
 combinatorial optimization problems:
  (a) multicriteria selection of the best change items,
  (b) selection of the best change items while taking into account
  the total resource requirement
 for the selected items (knapsack-like models);
  (c) selection of the best change items while taking
  into account
  the compatibility among the selected items
 (morphological clique problem).
 Note,
  special mathematical studies of discrete processes in evolution based on
  trees with application to phylogeny
 (including trees, metric on tree spaces, composition and
 decomposition of trees, compatibility of trees)
 are contained in
 \cite{steel16}.

 This article focuses on using the above-mentioned  author combinatorial approach
 to generations of wireless mobile communication systems
 (system modeling and changes).
 The described numerical illustrative examples
 (for generations of mobile communication system,
  for the system improvement process)
  can be used as a basis for
  the future modular system analysis (i.e., improvement, forecasting).

\newpage
\section{Generations of wireless communication systems}

 In recent years,
 several special research projects have payed  attention  to
 evolution of wireless communication systems
 (e.g.,
 \cite{bhala10a,fag14,gill15,hui03,kach14,kumars16,mehta14,sharma13,singh07,sur10,yadav17}).
 Evolution of wireless communications is based of
 communication standard generations.
 Basic evolution chain of
 wireless technology generations is presented in Fig. 1,
 a network-like evolution process is depicted in Fig. 2.

\begin{center}
\begin{picture}(134,78)
\put(17,00){\makebox(0,0)[bl]{Fig. 1. Basic evolution chain of
 wireless technology generations}}

\put(00,06){\vector(1,0){130}}
\put(130.5,05){\makebox(0,0)[bl]{t}}


\put(01,19){\makebox(0,0)[bl]{0G: clas-}}
\put(01,16){\makebox(0,0)[bl]{sical mo-}}
\put(01,13){\makebox(0,0)[bl]{bile te-}}
\put(01,10){\makebox(0,0)[bl]{lephony}}

\put(7.5,15.5){\oval(15,18)}

\put(12,23.5){\vector(3,1){05}}

\put(17,15){\line(1,0){15}} \put(17,30){\line(1,0){15}}
\put(17,15){\line(0,1){15}} \put(32,15){\line(0,1){15}}

\put(18,26){\makebox(0,0)[bl]{1G:}}
\put(18,23){\makebox(0,0)[bl]{analog }}
\put(18,20.5){\makebox(0,0)[bl]{cellular}}
\put(18,17){\makebox(0,0)[bl]{network}}

\put(18.5,011){\makebox(0,0)[bl]{\(\approx 1984\)}}

\put(27,30){\vector(3,1){07}}

\put(34,20){\line(1,0){15}} \put(34,35){\line(1,0){15}}
\put(34,20){\line(0,1){15}} \put(49,20){\line(0,1){15}}

\put(35,31){\makebox(0,0)[bl]{2G:}}
\put(35,27.6){\makebox(0,0)[bl]{digital }}
\put(35,25){\makebox(0,0)[bl]{wireless}}
\put(35,22){\makebox(0,0)[bl]{network}}

\put(33.5,16){\makebox(0,0)[bl]{\(1991..1992\)}}

\put(44,35){\vector(3,1){07}}

\put(51,25){\line(1,0){15}} \put(51,40){\line(1,0){15}}
\put(51,25){\line(0,1){15}} \put(66,25){\line(0,1){15}}

\put(52,36){\makebox(0,0)[bl]{3G:}}
\put(52,32.6){\makebox(0,0)[bl]{integra- }}
\put(52,30){\makebox(0,0)[bl]{tion of}}
\put(52,27){\makebox(0,0)[bl]{services}}

\put(50.5,21){\makebox(0,0)[bl]{\(2000...2001\)}}

\put(61,40){\vector(3,1){07}}

\put(68,30){\line(1,0){15}} \put(68,57){\line(1,0){15}}
\put(68,30){\line(0,1){27}} \put(83,30){\line(0,1){27}}

\put(69,53){\makebox(0,0)[bl]{4G:}}
\put(69,50){\makebox(0,0)[bl]{mobile }}
\put(69,47){\makebox(0,0)[bl]{commu-}}
\put(69,43.8){\makebox(0,0)[bl]{nication}}
\put(69,41.2){\makebox(0,0)[bl]{with}}
\put(69,38){\makebox(0,0)[bl]{higher}}
\put(69,35){\makebox(0,0)[bl]{require-}}
\put(69,32){\makebox(0,0)[bl]{ments}}

\put(67.5,26){\makebox(0,0)[bl]{\(2009...2010\)}}

\put(78,57){\vector(3,1){07}}

\put(85,46){\line(1,0){15}} \put(85,64){\line(1,0){15}}
\put(85,46){\line(0,1){18}} \put(100,46){\line(0,1){18}}

\put(86,60.4){\makebox(0,0)[bl]{5G:}}
\put(86,57){\makebox(0,0)[bl]{HetNets,}}
\put(86,54){\makebox(0,0)[bl]{high }}
\put(86,51){\makebox(0,0)[bl]{speed}}
\put(86,48){\makebox(0,0)[bl]{Internet}}

\put(86.5,42){\makebox(0,0)[bl]{\(\approx 2020\)}}

\put(95,64){\vector(3,1){07}}

\put(102,54){\line(1,0){15}} \put(102,69){\line(1,0){15}}
\put(102,54){\line(0,1){15}} \put(117,54){\line(0,1){15}}

\put(103,65){\makebox(0,0)[bl]{6G:}}
\put(103,62){\makebox(0,0)[bl]{5G \&}}
\put(103,59){\makebox(0,0)[bl]{satellite}}
\put(103,56){\makebox(0,0)[bl]{networks}}

\put(103.5,50){\makebox(0,0)[bl]{\(\approx 2025\)}}

\put(112,69){\vector(3,1){07}}

\put(119,56){\line(1,0){15}} \put(119,77){\line(1,0){15}}
\put(119,56){\line(0,1){21}} \put(134,56){\line(0,1){21}}

\put(120,73){\makebox(0,0)[bl]{7G:}}
\put(120,70){\makebox(0,0)[bl]{6G \&}}
\put(120,67){\makebox(0,0)[bl]{satellite}}
\put(119.7,64){\makebox(0,0)[bl]{functions}}
\put(120,60.8){\makebox(0,0)[bl]{(space }}
\put(120,58){\makebox(0,0)[bl]{roaming)}}

\end{picture}
\end{center}

\begin{center}
\begin{picture}(125,115)
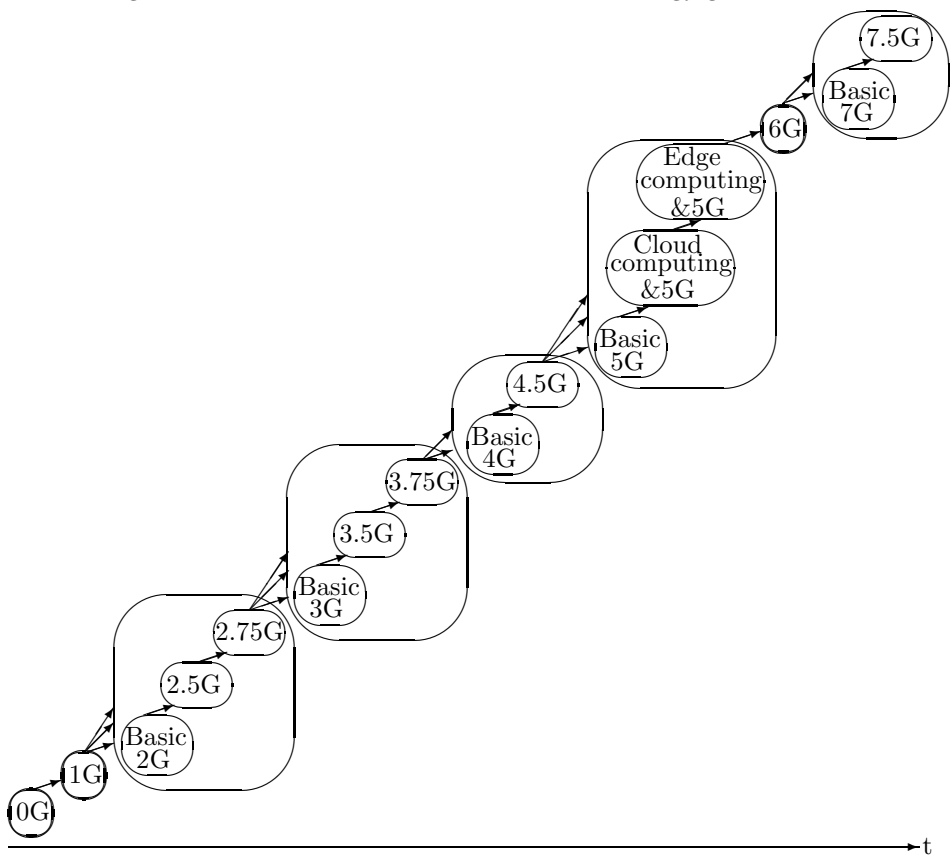
\put(05.5,00){\makebox(0,0)[bl]{Fig. 2. Simplified evolution
 network
  of wireless technology generations}}

\put(00,05.5){\vector(1,0){121}}
\put(121.5,04.5){\makebox(0,0)[bl]{t}}

\put(01,09){\makebox(0,0)[bl]{0G}} \put(03,10){\oval(06,06.5)}
\put(03,10){\oval(5.5,6)} \put(03,13){\vector(3,1){04}}

\put(08,14){\makebox(0,0)[bl]{1G}} \put(10,15){\oval(06,06.5)}
\put(10,15){\oval(5.5,6)} \put(10,18){\vector(3,1){04}}

\put(10,18){\vector(1,1){04}} \put(10,18){\vector(2,3){04}}

\put(26,26){\oval(24,26)}
\put(015.5,19){\makebox(0,0)[bl]{Basic}}
\put(017,16){\makebox(0,0)[bl]{2G}}

\put(19.75,19){\oval(09.5,08)} \put(18,23){\vector(3,1){04}}
\put(021,26){\makebox(0,0)[bl]{2.5G}} \put(25,27){\oval(09.5,06)}
\put(25,30){\vector(3,1){04}}
\put(027.5,33){\makebox(0,0)[bl]{2.75G}}
\put(32,34){\oval(09.5,06)}

\put(32,37){\vector(3,1){05.2}}

\put(32,37){\vector(1,1){05.2}}\put(32,37){\vector(2,3){05.2}}

\put(49,46){\oval(24,26)}
\put(38.5,39){\makebox(0,0)[bl]{Basic}}
\put(40,36){\makebox(0,0)[bl]{3G}}

\put(42.75,39){\oval(09.5,08)} \put(41,43){\vector(3,1){04}}
\put(44,46){\makebox(0,0)[bl]{3.5G}} \put(48,47){\oval(09.5,06)}
\put(48,50){\vector(3,1){04}}
\put(50.5,53){\makebox(0,0)[bl]{3.75G}}
\put(55,54){\oval(09.5,06)} \put(55,57){\vector(3,1){04}}

\put(55,57){\vector(1,1){04}}

\put(69,62.4){\oval(20,17)}
\put(61.5,59){\makebox(0,0)[bl]{Basic}}
\put(63,56){\makebox(0,0)[bl]{4G}}

\put(65.75,59){\oval(09.5,08)} \put(64,63){\vector(3,1){04}}
\put(67,66){\makebox(0,0)[bl]{4.5G}} \put(71,67){\oval(09.5,06)}

\put(71,70){\vector(3,1){06}}

\put(71,70){\vector(1,1){06}}\put(71,70){\vector(2,3){06}}

\put(89.5,83){\oval(25,33)}
\put(78.5,72){\makebox(0,0)[bl]{Basic}}
\put(80,69){\makebox(0,0)[bl]{5G}}

\put(82.75,72){\oval(09.5,08)} \put(81,76){\vector(3,1){04}}
\put(83,84.5){\makebox(0,0)[bl]{Cloud}}
\put(80,81.5){\makebox(0,0)[bl]{computing}}
\put(84,78.5){\makebox(0,0)[bl]{\&5G}}

\put(88,82.5){\oval(017,10)} \put(88,87.5){\vector(3,1){04}}
\put(87,95.5){\makebox(0,0)[bl]{Edge}}
\put(84,92.5){\makebox(0,0)[bl]{computing}}
\put(88,89.5){\makebox(0,0)[bl]{\&5G}}

\put(92,94){\oval(017,10)} \put(95,99){\vector(3,1){05}}

\put(101,100){\makebox(0,0)[bl]{6G}} \put(103,101){\oval(06,06.5)}
\put(103,101){\oval(5.5,6)} \put(103,104.5){\vector(3,1){04}}

\put(103,104.5){\vector(1,1){04}}

\put(116,108.2){\oval(18,17)}

\put(109,105){\makebox(0,0)[bl]{Basic}}
\put(110.5,102){\makebox(0,0)[bl]{7G}}

\put(113,105){\oval(09.5,08)} \put(111,109){\vector(3,1){04}}
\put(114,112){\makebox(0,0)[bl]{7.5G}}
\put(118,113){\oval(09.5,06)}

\end{picture}
\end{center}


 Table 5 contains
 descriptions of the generations:
 \(0G \rightarrow 1G \rightarrow  2G \rightarrow 3G \rightarrow
 4G \rightarrow 5G \rightarrow 6G \rightarrow 7G\).

\begin{center}
\begin{small}
 {\bf Table 5.} Generations of wireless mobile communication\\
\begin{tabular}{| c |  l | l |l |}
\hline

 No.& Generation &
  Brief description, features/attributes
 &Source(s) \\


\hline

 0.&0G technology (classical&(1) PTT (Push to Talk),&\cite{bhala10a,kumar10}\\
   &mobile telephony systems)&(2) MTS (Mobile telephone system),&\\
   &&(3) IMTS (Improved Mob. telephone system),&\\
   &&(4) AMTS (Adv. Mob. telephone system), etc.&\\

\hline
 1.&1G technology &AMPS (Analog Mobile Phone System) &\cite{bhala10a,gos15,sharma13}\\
   &(cellphones, analog wireless)
      &FDMA (Frequency Division Multiple Access)&\\

\hline
 2.&2G technology (digital wireless,& &\cite{bhala10a,gos15,mouly92}\\
 &  GPRS, EDGE)&& \cite{sharma13} \\

 2.1.&Basic 2G&GSM standard (circuit switched domain):&\cite{bhala10a,gos15,levvis07,liny97}\\
     &   &(a) TDMA (Time Division Multiple Access),&\cite{mehr97,mouly92,rah93}\\
        &&(b) CDMA (Code Division Multiple Access) &\\

 2.2.&2.5G GPRS (General Packet&Packet switched domain:&\cite{bhala10a,gos15}\\
   &  Radio Service) &(a) WAP, (b) MMS, (c) SMS &\\

 2.3.&2.75G EDGE (Enhanced Data&Extended version of GSM&\cite{bhala10a}\\
     &rates for GSM Evolution)&&\\

\hline
 3.&3G technology &Enhancements (over 2.5G):&\cite{bhala10a,chua00,gos15,hill02}\\
   &(Broad Band, IP Tech):&(1) enhancement of audio and video streaming,&\cite{kumara11,sharma13}\\
   &&(2) several times higher data speed,&\\
   &&(3) video-conferencing support,&\\
   &&(4) Web and WAP browsing at higher speed,&\\
   &&(5) IPTV (TV through the Internet) support.&\\

 3.1.&Basic 3G &Versions:&\cite{bhala10a,gos15,hill02}\\
       &&(1) W-CDMA, (2) GSM-EDGE, (3) UMTS,&\\

  &    &(4) DECT, (5) WiMAX, (6) CDMA 2000.&\\

 3.2.&3.5G-HSDPA (High-Speed Down- &For higher data transfer
 speed,
                 &\cite{bhala10a,gos15,peder06}\\
   &link  Packet  Access), W-CDMA&implementations: AMC, MIMO, HARQ&\\

 3.3.&3.75G-HSUPA (High-Speed & Enhancements: P2P data applications
     &\cite{bhala10a,kumars16}\\

   &Uplink Packet  Access), UMTS/ &(e.g., mobile e-mail, real-time P2P games),&\\
   &WCDMA  uplink evolution  &faster downloading Internet service, etc.&\\

\hline
 4.&4G technology:&&\\
 4.1.&Basic 4G (LTE, WiMAX, WiFi)&More bandwidth\&services offers; technolo-&\cite{bhala10a,boud09,bria01}\\
    & (enhanced WiMAX, 802.16, &gies: BDMA (Band Division Multiple Access),  &\cite{gho10,gos15,govil07,hui03}\\
    &3GPP, 3GPP2) (Internet),& CDMA; services: MMS, entertainment servi-  &\cite{kimy06,kumars16,kumara11}\\

    &service dynamic access&ces; Digital television in high definition, etc.&
       \cite{pari11,sharma13}\\

 4.2.&4.5G (LTE Advanced Pro) (\(4^{+}\)G) &Frequency bandwidth 1800 MHz  2100 MHz,
    & \cite{aky10,gho10,gho11}\\

   && mobile Internet speed 100 Mbit/sec&\\

\hline
5.&5G technologies:&&\\
 5.1.&Basic 5G technologies (IPv4):&New mobile revolution (high speed access)
   &\cite{abr16,akh16,andre13,andr14,bera16,bhala10a,bhu14,bocc14,bock16}\\

   &WLAN, 802.16 (WMAN), WPAN &(cell phones with very high bandwidth)&\cite{chin14,dahl14,deme13,ding15,gior16,gohil13,gos15}\\
   &(Internet), LTE-E (Long-Term& Services: OWA, OTP, Multimedia, & \cite{has14,imran14,jiang16,kach14,kimc16,kimk16} \\
   & Evolution-Enhanced),      &entertainment, radio  broadcasting,&\cite{mae16,mars16,mehta14,mich14,mitra15,mum14}\\
   & Heterogeneous networks (HetNet),& Digital Television, etc.&\cite{naik12,oss14,patil12,peng15,pir14}\\

   &asynchronous HetNet scenario& &\cite{rapp13,sharma13,sid11,singh12}\\
   && &\cite{thom14,tud11,tull16,zhou14}\\

 5.2.&Mob. cloud computing \&5G&&\cite{barb14,jom15,khan14,kwak15,you16}\\

 5.3.&Mob. edge computing (MEC)\&5G&3-tier heterogeneous MEC system
 &\cite{ahm16,mao17,munoz15,sabe16,tran16}\\

\hline
 6.&6G technology: 5G technology \&
  & Air-fiber technology, superfast broadband,
    &\cite{gos15,khu15,singh07}\\

 &satellite networks (telecommunica-   &etc.&\\
 &tion, Earth imaging, navigation)     &   &\\

\hline
 7.&7G technology: 6G technology \& & Satellite functions for mobile communication &\cite{khu15}\\
   &satellite communication functions  &  (space roaming) &\\

\hline
\end{tabular}
\end{small}
\end{center}

\section{Hierarchical structural models in communications}

\subsection{Examples of hierarchical multi-layer network
 architecture}

 Evidently, contemporary networks have multi-layer architecture/structure
 (i.e., functional/operational description, topology)
 \cite{current86,gavish92,kuz05,lev12hier,lev15,mur99,raza02,tanen02}.
 From the ``engineering'' viewpoint,
 the following hierarchical layers can be considered
  \cite{lev12hier,lev15,mur99,tanen02} (Fig. 3):

 {\bf 1.} Backbone network.

 {\bf 2.} Set of regional network clusters
 as interconnected network segments including the following:
 ~(a) additional centers (ie., hubs),
 ~(b) cross-connections, and ~(c) bridges.

 {\bf 3.} Access network/network segment (cluster):
 (e.g., bi-connected topology, about 20 nodes).

 {\bf 4.} Distributed network: a simple hard topology
 (e.g., bus, star, tree, ring).

 {\bf 5.} Layer of end-users (clients).

\begin{center}
\begin{picture}(101,98)

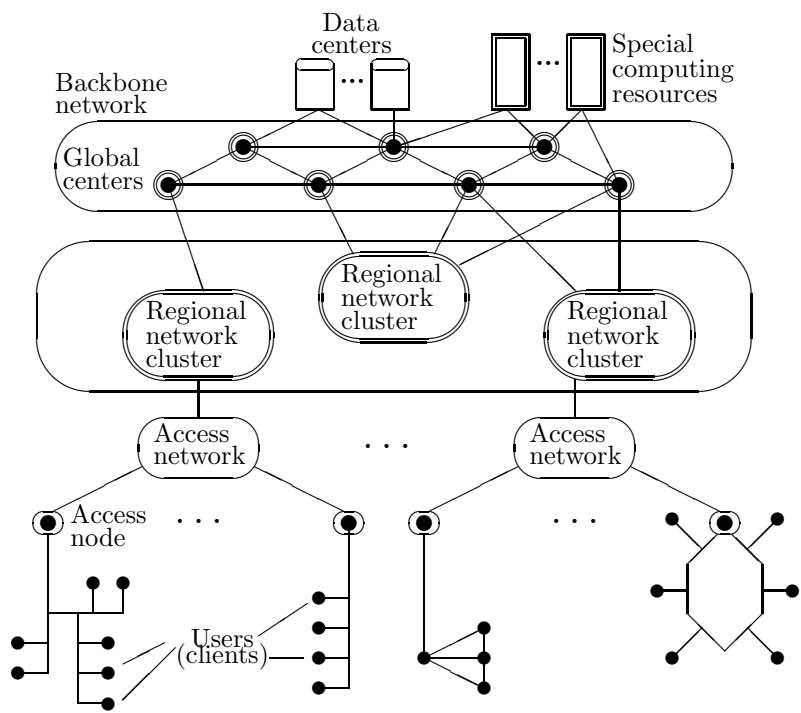
\put(09.5,00){\makebox(0,0)[bl]{Fig. 3. Example of
 multi-layer communication network}}


\put(65,85){\line(1,-1){05}} \put(65,85){\line(-3,-1){15}}

\put(75,85){\line(-1,-1){05}} \put(75,85){\line(1,-2){05}}

\put(63,85){\line(1,0){05}} \put(63,95){\line(1,0){5}}
\put(63,85){\line(0,1){10}} \put(68,85){\line(0,1){10}}

\put(63.5,85.5){\line(1,0){04}} \put(63.5,94.5){\line(1,0){4}}
\put(63.5,85.5){\line(0,1){09}} \put(67.5,85.5){\line(0,1){09}}

\put(68.8,91){\makebox(0,0)[bl]{{\bf ...}}}

\put(73,85){\line(1,0){05}} \put(73,95){\line(1,0){5}}
\put(73,85){\line(0,1){10}} \put(78,85){\line(0,1){10}}

\put(73.5,85.5){\line(1,0){04}} \put(73.5,94.5){\line(1,0){4}}
\put(73.5,85.5){\line(0,1){09}} \put(77.5,85.5){\line(0,1){09}}

\put(79,92){\makebox(0,0)[bl]{Special}}
\put(79,88.7){\makebox(0,0)[bl]{computing}}
\put(79,86){\makebox(0,0)[bl]{resources}}

\put(50,85){\line(0,-1){05}} \put(40,85){\line(-2,-1){10}}
\put(40,85){\line(2,-1){10}}

\put(39.5,91){\oval(5,1.5)}

\put(37,85){\line(1,0){05}}  \put(37,85){\line(0,1){06}}
\put(42,85){\line(0,1){06}}

\put(42.8,88.5){\makebox(0,0)[bl]{{\bf ...}}}

\put(49.5,91){\oval(5,1.5)}

\put(47,85){\line(1,0){05}} \put(47,85){\line(0,1){06}}
\put(52,85){\line(0,1){06}}

\put(40.5,95.5){\makebox(0,0)[bl]{Data }}
\put(39,92.5){\makebox(0,0)[bl]{centers}}

\put(50,77.5){\oval(90,12)}

\put(05,87.5){\makebox(0,0)[bl]{Backbone}}
\put(05,84.5){\makebox(0,0)[bl]{network}}

\put(06,77.5){\makebox(0,0)[bl]{Global}}
\put(06,74.5){\makebox(0,0)[bl]{centers}}


\put(80,75){\circle{4.0}} \put(80,75){\circle{3.2}}
\put(80,75){\circle*{2}} \put(80,75){\line(-1,0){20}}
\put(80,75){\line(-2,-1){21.3}} \put(80,75){\line(0,-1){14.2}}

\put(60,75){\circle{4.0}} \put(60,75){\circle{3.2}}
\put(60,75){\circle*{2}}

\put(60,75){\line(-1,-2){04.6}} \put(60,75){\line(1,-1){014.2}}

\put(40,75){\circle{4.0}} \put(40,75){\circle{3.2}}
\put(40,75){\circle*{2}}

\put(40,75){\line(1,0){20}} \put(40,75){\line(1,-2){04.6}}

\put(20,75){\circle{4.0}} \put(20,75){\circle{3.2}}
\put(20,75){\circle*{2}} \put(20,75){\line(1,0){20}}

\put(20,75){\line(1,-3){04.7}}

\put(30,80){\circle{4.0}} \put(30,80){\circle{3.2}}
\put(30,80){\circle*{2}} \put(30,80){\line(-2,-1){10}}
\put(30,80){\line(2,-1){10}}

\put(50,80){\circle{4.0}} \put(50,80){\circle{3.2}}
\put(50,80){\circle*{2}} \put(50,80){\line(1,0){20}}
\put(50,80){\line(-1,0){20}} \put(50,80){\line(-2,-1){10}}
\put(50,80){\line(2,-1){10}}

\put(70,80){\circle{4.0}} \put(70,80){\circle{3.2}}
\put(70,80){\circle*{2}} \put(70,80){\line(-2,-1){10}}
\put(70,80){\line(2,-1){10}}

\put(50,57.5){\oval(95,20)}


\put(43,61.4){\makebox(0,0)[bl]{Regional}}
\put(43,58.8){\makebox(0,0)[bl]{network}}
\put(43,55.8){\makebox(0,0)[bl]{cluster}}

\put(50,60){\oval(20,12)} \put(50,60){\oval(19,11)}

\put(73,56.4){\makebox(0,0)[bl]{Regional}}
\put(73,53.8){\makebox(0,0)[bl]{network}}
\put(73,50.8){\makebox(0,0)[bl]{cluster}}

\put(80,55){\oval(20,12)} \put(80,55){\oval(19,11)}

\put(17,56.4){\makebox(0,0)[bl]{Regional}}
\put(17,53.8){\makebox(0,0)[bl]{network}}
\put(17,50.8){\makebox(0,0)[bl]{cluster}}

\put(24,55){\oval(20,12)} \put(24,55){\oval(19,11)}


\put(46,40){\makebox(0,0)[bl]{{\bf .~.~.}}}

\put(18,40.8){\makebox(0,0)[bl]{Access}}
\put(18,37.8){\makebox(0,0)[bl]{network}}

\put(24,44){\line(0,1){05}}

\put(24,40){\oval(16,8)}

\put(21,30){\makebox(0,0)[bl]{{\bf .~.~.}}}

\put(23,13.5){\makebox(0,0)[bl]{Users}}
\put(21,10.5){\makebox(0,0)[bl]{(clients)}}

\put(19.5,14){\line(-2,-1){5.5}} \put(21,13.5){\line(-1,-1){7.1}}

\put(32,15){\line(3,2){6.5}} \put(33.5,12){\line(1,0){4.5}}


\put(43.5,31.5){\line(-2,1){12}}

\put(44,30){\oval(4,3)} \put(44,30){\circle*{2}}


\put(44,30){\line(0,-1){22}}

\put(40,08){\circle*{1.7}} \put(40,08){\line(1,0){4}}
\put(40,20){\circle*{1.7}} \put(40,20){\line(1,0){4}}
\put(40,16){\circle*{1.7}} \put(40,16){\line(1,0){4}}
\put(40,12){\circle*{1.7}} \put(40,12){\line(1,0){4}}


\put(07,30){\makebox(0,0)[bl]{Access}}
\put(07,27){\makebox(0,0)[bl]{node}}

\put(04.5,31.5){\line(2,1){12}}

\put(04,30){\oval(4,3)} \put(04,30){\circle*{2}}


\put(00,10){\circle*{1.7}} \put(00,14){\circle*{1.7}}
\put(00,10){\line(1,0){04}} \put(00,14){\line(1,0){04}}
\put(04,10){\line(0,1){18.5}}


\put(12,10){\circle*{1.7}}  \put(12,14){\circle*{1.7}}
\put(12,06){\circle*{1.7}}

\put(12,06){\line(-1,0){04}} \put(12,10){\line(-1,0){04}}
\put(12,14){\line(-1,0){04}}

\put(08,06){\line(0,1){12}}


\put(04,18){\line(1,0){10}}

\put(10,22){\circle*{1.7}}  \put(14,22){\circle*{1.7}}
\put(10,18){\line(0,1){04}} \put(14,18){\line(0,1){04}}



\put(68,40.8){\makebox(0,0)[bl]{Access}}
\put(68,37.8){\makebox(0,0)[bl]{network}}

\put(74,44){\line(0,1){05}}

\put(74,40){\oval(16,8)}

\put(71,30){\makebox(0,0)[bl]{{\bf .~.~.}}}


\put(93.5,31.5){\line(-2,1){12}}

\put(94,30){\oval(4,3)} \put(94,30){\circle*{2}}

\put(89,24.5){\line(1,1){4}} \put(99,24.5){\line(-1,1){4}}

\put(87,30.5){\circle*{1.7}} \put(101,30.5){\circle*{1.7}}
\put(91,26.5){\line(-1,1){04}} \put(97,26.5){\line(1,1){04}}

\put(89,18){\line(0,1){6.5}} \put(99,18){\line(0,1){6.5}}

\put(85,21){\circle*{1.7}} \put(103,21){\circle*{1.7}}
\put(89,21){\line(-1,0){04}} \put(99,21){\line(1,0){04}}

\put(89,18){\line(1,-1){5}} \put(99,18){\line(-1,-1){5}}

\put(87,12){\circle*{1.7}} \put(101,12){\circle*{1.7}}
\put(91,16){\line(-1,-1){04}} \put(97,16){\line(1,-1){04}}


\put(54.5,31.5){\line(2,1){12}}

\put(54,30){\oval(4,3)} \put(54,30){\circle*{2}}

\put(54,12){\circle*{1.7}}

\put(54,12){\line(0,1){16.5}}

\put(54,12){\line(1,0){08}} \put(54,12){\line(2,1){08}}
\put(54,12){\line(2,-1){08}} \put(62,08){\line(0,1){08}}

\put(62,12){\circle*{1.7}}  \put(62,16){\circle*{1.7}}
\put(62,08){\circle*{1.7}}

\end{picture}
\end{center}

  A three-layer network topology based on central hub subnetwork
 is depicted in Fig. 4:
 (1) layer of terminal nodes (clients/end users),
 (2) layer of hubs and/or composite hubs, and
 (3) layer of subnetwork of central hubs.

\begin{center}
\begin{picture}(90,57)
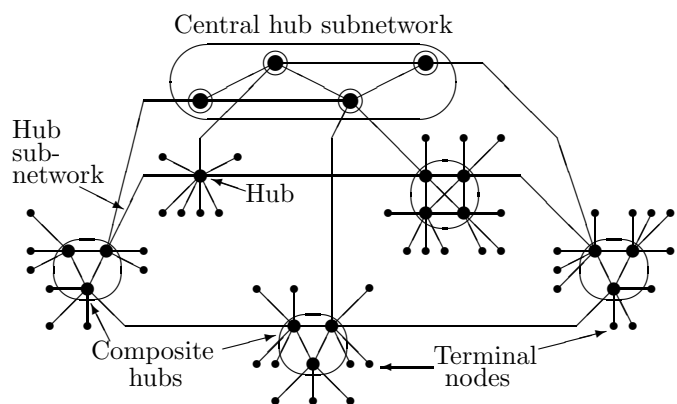
\put(012,00){\makebox(0,0)[bl]{Fig. 4. Three-layer network
  architecture}}

\put(21.5,54){\makebox(0,0)[bl]{Central hub subnetwork}}

\put(00,40){\makebox(0,0)[bl]{Hub}}
\put(00,37){\makebox(0,0)[bl]{sub-}}
\put(00,34){\makebox(0,0)[bl]{network}}
\put(09,33.5){\vector(2,-1){05.8}}

\put(40,47.5){\oval(38,010)}

\put(35,50){\circle{3.1}} \put(35,50){\circle*{2.1}}

\put(55,50){\circle{3.1}} \put(55,50){\circle*{2.1}}

\put(45,45){\circle{3.1}} \put(45,45){\circle*{2.1}}

\put(25,45){\circle{3.1}} \put(25,45){\circle*{2.1}}

\put(25,45){\line(1,0){20}} \put(25,45){\line(2,1){10}}

\put(45,45){\line(2,1){10}} \put(45,45){\line(-2,1){10}}
\put(35,50){\line(1,0){20}}

\put(31,31.5){\makebox(0,0)[bl]{Hub}}
\put(31,33){\vector(-3,1){04.5}}

\put(25,35){\line(0,1){05}} \put(25,40){\line(1,1){10}}

\put(25,35){\circle*{1.6}}
\put(22.5,30){\circle*{1.0}} \put(22.5,30){\line(1,2){02.5}}
\put(25,30){\circle*{1.0}} \put(25,30){\line(0,1){05}}
\put(27.5,30){\circle*{1.0}} \put(27.5,30){\line(-1,2){2.5}}

\put(20,30){\circle*{1.0}} \put(20,30){\line(1,1){05}}

\put(20,37.5){\circle*{1.0}} \put(20,37.5){\line(2,-1){05}}
\put(30,37.5){\circle*{1.0}} \put(30,37.5){\line(-2,-1){05}}

\put(55,35){\line(-1,1){10}}

\put(57.5,32.5){\oval(09,09)}

\put(55,35){\circle*{1.6}} \put(60,35){\circle*{1.6}}
\put(55,30){\circle*{1.6}} \put(60,30){\circle*{1.6}}

\put(55,30){\line(1,0){05}} \put(55,35){\line(1,0){05}}
\put(55,30){\line(0,1){05}} \put(60,30){\line(0,1){05}}
\put(55,30){\line(1,1){05}} \put(60,30){\line(-1,1){05}}

\put(52.5,25){\circle*{1.0}} \put(52.5,25){\line(1,2){02.5}}
\put(55,25){\circle*{1.0}} \put(55,25){\line(0,1){05}}
\put(57.5,25){\circle*{1.0}} \put(57.5,25){\line(-1,2){02.5}}

\put(65,30){\circle*{1.0}} \put(65,30){\line(-1,0){05}}
\put(62.5,25){\circle*{1.0}} \put(62.5,25){\line(-1,2){02.5}}

\put(55,40){\circle*{1.0}} \put(55,40){\line(0,-1){05}}
\put(60,40){\circle*{1.0}} \put(60,40){\line(0,-1){05}}
\put(65,40){\circle*{1.0}} \put(65,40){\line(-1,-1){05}}
\put(50,30){\circle*{1.0}} \put(50,30){\line(1,0){05}}
\put(65,25){\circle*{1.0}} \put(65,25){\line(-1,1){05}}

\put(60,35){\line(1,0){07.5}} \put(67.5,35){\line(1,-1){10}}

\put(20,35){\line(1,0){35}}

\put(12.5,25){\line(1,2){05}} \put(17.5,35){\line(1,0){07.5}}
\put(37.5,15){\line(-1,0){22.5}} \put(15,15){\line(-1,1){05}}
\put(42.5,15){\line(1,0){32.5}} \put(75,15){\line(1,1){05}}

\put(42.5,15){\line(0,1){25}} \put(42.5,40){\line(1,2){02.5}}

\put(10.5,10){\makebox(0,0)[bl]{Composite}}
\put(15.5,07){\makebox(0,0)[bl]{hubs}}
\put(13,13){\vector(-1,2){02.5}} \put(28,12){\vector(4,1){07}}

\put(40,12.5){\oval(09,08)}

\put(37.5,15){\circle*{1.6}} \put(42.5,15){\circle*{1.6}}
\put(40,10){\circle*{1.6}}

\put(40,10){\line(1,2){02.5}} \put(40,10){\line(-1,2){02.5}}
\put(37.5,15){\line(1,0){05}}

\put(35,05){\circle*{1.0}} \put(35,05){\line(1,1){05}}
\put(40,05){\circle*{1.0}} \put(40,05){\line(0,1){05}}
\put(45,05){\circle*{1.0}} \put(45,05){\line(-1,1){05}}

\put(45,10){\circle*{1.0}} \put(45,10){\line(-1,2){02.5}}
\put(35,10){\circle*{1.0}} \put(35,10){\line(1,2){02.5}}

\put(32.5,10){\circle*{1.0}} \put(32.5,10){\line(1,1){05}}
\put(47.5,10){\circle*{1.0}} \put(47.5,10){\line(-1,1){05}}

\put(37.5,20){\circle*{1.0}} \put(37.5,20){\line(0,-1){05}}
\put(32.5,20){\circle*{1.0}} \put(32.5,20){\line(1,-1){05}}
\put(47.5,20){\circle*{1.0}} \put(47.5,20){\line(-1,-1){05}}

\put(56,10){\makebox(0,0)[bl]{Terminal}}
\put(57.5,07){\makebox(0,0)[bl]{nodes}}
\put(56,09.5){\vector(-1,0){07}} \put(70,12){\vector(4,1){09}}


\put(12.5,25){\line(1,4){05}} \put(17.5,45){\line(1,0){7.5}}


\put(10,22.5){\oval(09,08)}

\put(07.5,25){\circle*{1.6}} \put(12.5,25){\circle*{1.6}}
\put(10,20){\circle*{1.6}}

\put(10,20){\line(1,2){02.5}} \put(10,20){\line(-1,2){02.5}}
\put(07.5,25){\line(1,0){05}}

\put(05,15){\circle*{1.0}} \put(05,15){\line(1,1){05}}
\put(10,15){\circle*{1.0}} \put(10,15){\line(0,1){05}}

\put(17.5,22.5){\circle*{1.0}} \put(17.5,22.5){\line(-2,1){05}}
\put(17.5,25){\circle*{1.0}} \put(17.5,25){\line(-1,0){05}}

\put(02.5,25){\circle*{1.0}} \put(02.5,25){\line(1,0){05}}
\put(02.5,22.5){\circle*{1.0}} \put(02.5,22.5){\line(2,1){05}}

\put(02.5,30){\circle*{1.0}} \put(02.5,30){\line(1,-1){05}}
\put(05,20){\circle*{1.0}} \put(05,20){\line(1,0){05}}

\put(77.5,25){\line(-1,3){05}} \put(72.5,40){\line(-1,1){10}}
\put(62.5,50){\line(-1,0){07.5}}

\put(80,22.5){\oval(09,08)}

\put(77.5,25){\circle*{1.6}} \put(82.5,25){\circle*{1.6}}
\put(80,20){\circle*{1.6}}

\put(80,20){\line(1,2){02.5}} \put(80,20){\line(-1,2){02.5}}
\put(77.5,25){\line(1,0){05}}

\put(82.5,15){\circle*{1.0}} \put(82.5,15){\line(-1,2){02.5}}
\put(80,15){\circle*{1.0}} \put(80,15){\line(0,1){05}}

\put(85,30){\circle*{1.0}} \put(85,30){\line(-1,-2){02.5}}
\put(87.5,30){\circle*{1.0}} \put(87.5,30){\line(-1,-1){05}}

\put(87.5,25){\circle*{1.0}} \put(87.5,25){\line(-1,0){05}}

\put(72.5,20){\circle*{1.0}} \put(72.5,20){\line(1,1){05}}
\put(72.5,22.5){\circle*{1.0}} \put(72.5,22.5){\line(2,1){05}}

\put(85,20){\circle*{1.0}} \put(85,20){\line(-1,0){05}}
\put(82.5,30){\circle*{1.0}} \put(82.5,30){\line(0,-1){05}}
\put(77.5,30){\circle*{1.0}} \put(77.5,30){\line(0,-1){05}}

\put(72.5,25){\circle*{1.0}} \put(72.5,25){\line(1,0){05}}

\end{picture}
\end{center}

 An illustrative scheme of functional multi-layer architecture for 5G system
 is depicted in Fig. 5
  \cite{luww08}.

\begin{center}
\begin{picture}(100,62)
\put(11.5,00){\makebox(0,0)[bl] {Fig. 5.
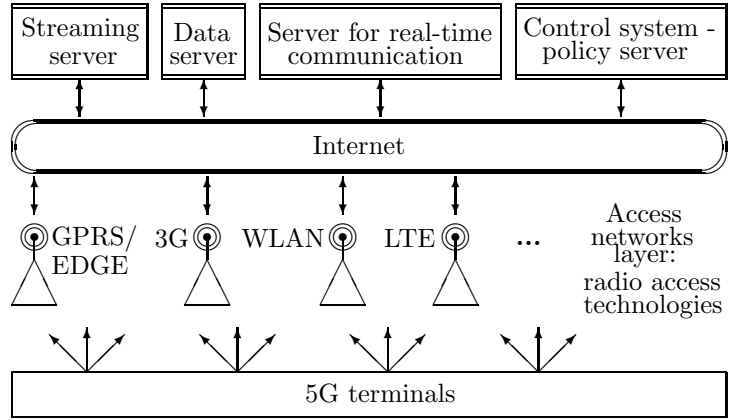
  Functional architecture of 5G system}}

\put(01.3,55){\makebox(0,0)[bl]{Streaming}}
\put(05,52){\makebox(0,0)[bl]{server}}

\put(00,50.5){\line(1,0){18}} \put(00,59.5){\line(1,0){18}}
\put(00,50){\line(1,0){18}} \put(00,60){\line(1,0){18}}
\put(00,50){\line(0,1){10}} \put(18,50){\line(0,1){10}}

\put(21.5,55){\makebox(0,0)[bl]{Data}}
\put(21,52){\makebox(0,0)[bl]{server}}

\put(20,50.5){\line(1,0){11}} \put(20,59.5){\line(1,0){11}}
\put(20,50){\line(1,0){11}} \put(20,60){\line(1,0){11}}
\put(20,50){\line(0,1){10}} \put(31,50){\line(0,1){10}}

\put(34,55){\makebox(0,0)[bl]{Server for real-time}}
\put(36.7,52){\makebox(0,0)[bl]{communication}}

\put(33,50.5){\line(1,0){32}} \put(33,59.5){\line(1,0){32}}
\put(33,50){\line(1,0){32}} \put(33,60){\line(1,0){32}}
\put(33,50){\line(0,1){10}} \put(65,50){\line(0,1){10}}

\put(68,55){\makebox(0,0)[bl]{Control system -}}
\put(70.5,52){\makebox(0,0)[bl]{policy server}}

\put(67,50.5){\line(1,0){28}} \put(67,59.5){\line(1,0){28}}
\put(67,50){\line(1,0){28}} \put(67,60){\line(1,0){28}}
\put(67,50){\line(0,1){10}} \put(95,50){\line(0,1){10}}

\put(09,46){\vector(0,1){4}} \put(09,49){\vector(0,-1){4}}
\put(25.5,46){\vector(0,1){4}} \put(25.5,49){\vector(0,-1){4}}
\put(49,46){\vector(0,1){4}} \put(49,49){\vector(0,-1){4}}
\put(81,46){\vector(0,1){4}} \put(81,49){\vector(0,-1){4}}

\put(40,40){\makebox(0,0)[bl]{Internet}}
\put(47.5,41){\oval(95,07)} \put(47.5,41){\oval(94,06.3)}

\put(03,33){\vector(0,1){4}} \put(03,36){\vector(0,-1){4}}
\put(26,33){\vector(0,1){4}} \put(26,36){\vector(0,-1){4}}
\put(44,33){\vector(0,1){4}} \put(44,36){\vector(0,-1){4}}
\put(59,33){\vector(0,1){4}} \put(59,36){\vector(0,-1){4}}

\put(39,07){\makebox(0,0)[bl]{5G terminals}}

\put(00,05){\line(1,0){95}} \put(00,11){\line(1,0){95}}
\put(00,05){\line(0,1){6}} \put(95,05){\line(0,1){6}}

\put(05.3,27){\makebox(0,0)[bl]{GPRS/}}
\put(05.3,24){\makebox(0,0)[bl]{EDGE}}

\put(00,20){\line(1,0){6}} \put(00,20){\line(1,2){3}}
\put(06,20){\line(-1,2){3}} \put(03,26){\line(0,1){3}}
\put(03,29){\circle*{1}}

\put(03,29){\circle{2.5}} \put(03,29){\circle{3.5}}
\put(10,11){\vector(0,1){6}} \put(10,11){\vector(-1,1){5}}
\put(10,11){\vector(1,1){5}}

\put(30,11){\vector(0,1){6}} \put(30,11){\vector(-1,1){5}}
\put(30,11){\vector(1,1){5}}

\put(50,11){\vector(0,1){6}} \put(50,11){\vector(-1,1){5}}
\put(50,11){\vector(1,1){5}}

\put(70,11){\vector(0,1){6}} \put(70,11){\vector(-1,1){5}}
\put(70,11){\vector(1,1){5}}

\put(19,27.6){\makebox(0,0)[bl]{3G}}

\put(23,20){\line(1,0){6}} \put(23,20){\line(1,2){3}}
\put(29,20){\line(-1,2){3}} \put(26,26){\line(0,1){3}}
\put(26,29){\circle*{1}}

\put(26,29){\circle{2.5}} \put(26,29){\circle{3.5}}

\put(30.5,27.6){\makebox(0,0)[bl]{WLAN}}

\put(41,20){\line(1,0){6}} \put(41,20){\line(1,2){3}}
\put(47,20){\line(-1,2){3}} \put(44,26){\line(0,1){3}}
\put(44,29){\circle*{1}}

\put(44,29){\circle{2.5}} \put(44,29){\circle{3.5}}

\put(49.5,27.6){\makebox(0,0)[bl]{LTE}}

\put(56,20){\line(1,0){6}} \put(56,20){\line(1,2){3}}
\put(62,20){\line(-1,2){3}} \put(59,26){\line(0,1){3}}
\put(59,29){\circle*{1}}

\put(59,29){\circle{2.5}} \put(59,29){\circle{3.5}}

\put(67,28){\makebox(0,0)[bl]{{\bf...}}}

\put(79,31){\makebox(0,0)[bl]{Access  }}
\put(77,28){\makebox(0,0)[bl]{networks }}
\put(80,25){\makebox(0,0)[bl]{layer: }}
\put(76,22){\makebox(0,0)[bl]{radio access }}
\put(76,18.2){\makebox(0,0)[bl]{technologies}}

\end{picture}
\end{center}


 Table 6 contains a description of
  hierarchical six-layer architecture of space information networks (SINs)
 \cite{gou17,luy13,quan16,zhangw17}.

\begin{center}
 {\bf Table 6.} Hierarchy of space information networks
  (SINs) \cite{gou17,luy13,quan16,zhangw17} \\
\begin{tabular}{| c | l | l |}
\hline
 No.&Layer & Networks/Systems \\

\hline

  I. & Satellite layers:&GEO backbone networks:\\
     &&(i) distributed satellite clusters DSCs,\\
     &&(ii) service enhanced satellite networks,\\
     &&(iii) navigation satellites\\

 1.1.&Geostationary orbits (high 36 000 km)&'Central network hubs'\\

   &(geostationary satellites)&\\

 1.2.&Medium earth orbits (high 30 000 km)&\\

    &(medium orbit satellites)&\\

 1.3.&Low earth orbits (low-orbit satellites)&'Hubs' for lower-level networks\\


       && deep-space explorers\\

 II.& Near space (600 km)
             &(1) high attitude platform networks \\
  &&(HAPs),\\

   &&(2)  near-earth orbit earth observation \\
     &&  satellites\\

 III.&Stratospheric layer (20 km):&stratospheric airships network\\

 IV.& Air (e.g., airplanes) (10 km) & airplanes based networks \\

 V.& Low-attitude terminals (e.g., balloons) (\(<\) 1000 m) &balloon networks \\

 VI.& Ground systems and nodes (0 km)&Systems:\\
  && (a) operation and control networks,\\
  &&(b) access networks,\\
  &&(c) ground information grid networks,\\
  &&(d) information processing centers,\\
  &&(e) air traffic control centers,\\
  &&(f) community region networks,\\
  &&(g) ground users.\\

\hline
\end{tabular}
\end{center}

\newpage
\subsection{General hierarchical structural models of communication
 technology}

 Generally, seven-layer Open System Interconnection model (OSI model)
 is considered as a basic conceptual model for
 telecommunication and computer systems \cite{miao14,zimm80}:


 {\bf Layer 1.} Physical layer (\(L^{1}\)):~
 {\it Protocol data unit:} Bit.
 {\it Functions:} Transmission and reception of raw bit streams
 over a physical medium.

 {\bf Layer 2.} Data link layer (\(L^{2}\)):~
 {\it Protocol data unit:} Frame.
 {\it Functions:} Reliable transmission
 of data streams between two nodes connected by a physical layer.

  {\bf Layer 3.} Network layer (\(L^{3}\)):~
 {\it Protocol data unit:} Packet.
 {\it Functions:} Structuring and managing a multi-node network,
 including addressing,
 routing, and traffic control.

 {\bf  Layer 4.} Transport layer (\(L^{4}\)):~
 {\it Protocol data unit:} Segment, Datagram.
 {\it Functions:} Reliable transmission of data segments between points
 on a network
 (including segmentation, acknowledgement
 and multiplexing).

 {\bf Layer 5.} Session layer (\(L^{5}\)):~
 {\it Protocol data unit:} Data.
 {\it Functions:} Managing communication sessions
 (i.e., controlling the dialogs (connections) between computers; exchange of information
 between two nodes).

  {\bf Layer 6.} Presentation layer (\(L^{6}\)):~
 {\it Protocol data unit:} Data.
 {\it Functions:} Translation of data between a networking service and
 an application.

 {\bf Layer 7.} Application layer (\(L^{7}\)):~
 {\it Protocol data unit:} Data.
 {\it Functions:} resource charing, remote access file access, etc.


 Thus, the structure of the communication system \(S\)
  can be presented as the corresponding  seven-part morphological hierarchy
 (Fig. 6),
 where
  for each layer \(\kappa = \overline{1,7}\)
 alternative implementation versions are:
 \(L^{\kappa}_{1},...,L^{\kappa}_{\iota_{\kappa}},...,L^{\kappa}_{q_{\kappa}}\).

\begin{center}
\begin{picture}(100,46)
\put(00,00){\makebox(0,0)[bl] {Fig. 6.
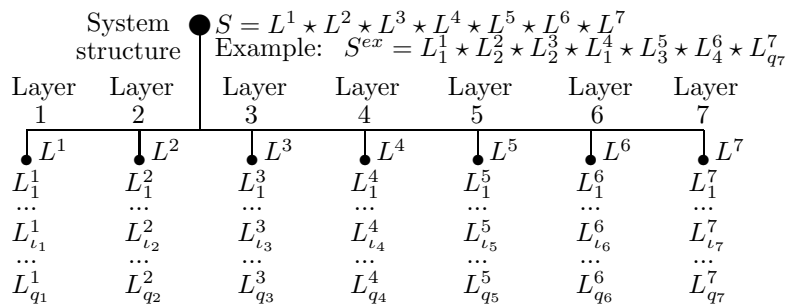
  Seven layer based structure of communication system}}

\put(25,42){\circle*{2.7}}

\put(27,41){\makebox(0,0)[bl]{\(S =
  L^{1}\star L^{2}\star L^{3}\star L^{4}
 \star L^{5} \star L^{6}\star L^{7} \)}}

\put(27,36.5){\makebox(0,0)[bl]{Example: ~\(S^{ex} =
 L^{1}_{1} \star L^{2}_{2} \star L^{3}_{2}
 \star L^{4}_{1}
 \star L^{5}_{3} \star L^{6}_{4} \star L^{7}_{q_{7}}
 \)}}

\put(10,40.5){\makebox(0,0)[bl]{System}}
\put(08.6,37.5){\makebox(0,0)[bl]{structure}}

\put(25,42){\line(0,-1){14}}

\put(02,28){\line(1,0){90}}

\put(00,32){\makebox(0,0)[bl]{Layer}}
\put(03,29){\makebox(0,0)[bl]{1}}

\put(02,24){\line(0,1){4}} \put(02,24){\circle*{1.4}}

\put(03.5,24){\makebox(0,0)[bl]{\(L^{1}\)}}

\put(00,19){\makebox(0,0)[bl]{\(L^{1}_{1}\)}}
\put(0.5,17.4){\makebox(0,0)[bl]{...}}
\put(00,12){\makebox(0,0)[bl]{\(L^{1}_{\iota_{1}}\) }}
\put(0.5,10.4){\makebox(0,0)[bl]{...}}
\put(00,05){\makebox(0,0)[bl]{\(L^{1}_{q_{1}}\)}}

\put(13,32){\makebox(0,0)[bl]{Layer}}
\put(16,29){\makebox(0,0)[bl]{2}}

\put(17,24){\line(0,1){4}} \put(17,24){\circle*{1.4}}

\put(18.5,24){\makebox(0,0)[bl]{\(L^{2}\)}}

\put(15,19){\makebox(0,0)[bl]{\(L^{2}_{1}\)}}
\put(15.5,17.4){\makebox(0,0)[bl]{...}}
\put(15,12){\makebox(0,0)[bl]{\(L^{2}_{\iota_{2}}\) }}
\put(15.5,10.4){\makebox(0,0)[bl]{...}}
\put(15,05){\makebox(0,0)[bl]{\(L^{2}_{q_{2}}\)}}

\put(28,32){\makebox(0,0)[bl]{Layer}}
\put(31,29){\makebox(0,0)[bl]{3}}

\put(32,24){\line(0,1){4}} \put(32,24){\circle*{1.4}}

\put(33.5,24){\makebox(0,0)[bl]{\(L^{3}\)}}

\put(30,19){\makebox(0,0)[bl]{\(L^{3}_{1}\)}}
\put(30.5,17.4){\makebox(0,0)[bl]{...}}
\put(30,12){\makebox(0,0)[bl]{\(L^{3}_{\iota_{3}}\) }}
\put(30.5,10.4){\makebox(0,0)[bl]{...}}
\put(30,05){\makebox(0,0)[bl]{\(L^{3}_{q_{3}}\)}}

\put(43,32){\makebox(0,0)[bl]{Layer}}
\put(46,29){\makebox(0,0)[bl]{4}}

\put(47,24){\line(0,1){4}} \put(47,24){\circle*{1.4}}

\put(48.5,24){\makebox(0,0)[bl]{\(L^{4}\)}}

\put(45,19){\makebox(0,0)[bl]{\(L^{4}_{1}\)}}
\put(45.5,17.4){\makebox(0,0)[bl]{...}}
\put(45,12){\makebox(0,0)[bl]{\(L^{4}_{\iota_{4}}\) }}
\put(45.5,10.4){\makebox(0,0)[bl]{...}}
\put(45,05){\makebox(0,0)[bl]{\(L^{4}_{q_{4}}\)}}

\put(58,32){\makebox(0,0)[bl]{Layer}}
\put(61,29){\makebox(0,0)[bl]{5}}

\put(62,24){\line(0,1){4}} \put(62,24){\circle*{1.4}}

\put(63.5,24){\makebox(0,0)[bl]{\(L^{5}\)}}

\put(60,19){\makebox(0,0)[bl]{\(L^{5}_{1}\)}}
\put(60.5,17.4){\makebox(0,0)[bl]{...}}
\put(60,12){\makebox(0,0)[bl]{\(L^{5}_{\iota_{5}}\) }}
\put(60.5,10.4){\makebox(0,0)[bl]{...}}
\put(60,05){\makebox(0,0)[bl]{\(L^{5}_{q_{5}}\)}}

\put(74,32){\makebox(0,0)[bl]{Layer}}
\put(77,29){\makebox(0,0)[bl]{6}}

\put(77,24){\line(0,1){4}} \put(77,24){\circle*{1.4}}

\put(78.5,24){\makebox(0,0)[bl]{\(L^{6}\)}}

\put(75,19){\makebox(0,0)[bl]{\(L^{6}_{1}\)}}
\put(75.5,17.4){\makebox(0,0)[bl]{...}}
\put(75,12){\makebox(0,0)[bl]{\(L^{6}_{\iota_{6}}\) }}
\put(75.5,10.4){\makebox(0,0)[bl]{...}}
\put(75,05){\makebox(0,0)[bl]{\(L^{6}_{q_{6}}\)}}

\put(88,32){\makebox(0,0)[bl]{Layer}}
\put(91,29){\makebox(0,0)[bl]{7}}

\put(92,24){\line(0,1){4}} \put(92,24){\circle*{1.4}}

\put(93.5,24){\makebox(0,0)[bl]{\(L^{7}\)}}

\put(90,19){\makebox(0,0)[bl]{\(L^{7}_{1}\)}}
\put(90.5,17.4){\makebox(0,0)[bl]{...}}
\put(90,12){\makebox(0,0)[bl]{\(L^{7}_{\iota_{7}}\) }}
\put(90.5,10.4){\makebox(0,0)[bl]{...}}
\put(90,05){\makebox(0,0)[bl]{\(L^{7}_{q_{7}}\)}}

\end{picture}
\end{center}

 Another hierarchical internetworking model
 for enterprise networks
 (a three-layer model for network design)
 has been proposed by Cisco
 \cite{raza02} (Fig. 7):


 {\bf  1. Access layer}:
 connecting client nodes (e.g., workstations) to the network.

 {\bf 2. Distribution layer} (work group layer):
 management of  routing, filtering, and QoS policies,
 management of individual branch-office WAN connections.

 {\bf 3. Core layer} (core network):
 high-speed, highly redundant forwarding services to move
 packets between distribution-layer devices in different regions of
 the network
 (core network devices manage the highest-speed connections).

\begin{center}
\begin{picture}(94,44)
\put(10,00){\makebox(0,0)[bl]{Fig. 7.
 Three-layer model for enterprise network}}

\put(027,38){\makebox(0,0)[bl]{Core layer (core network):}}
\put(06.6,35){\makebox(0,0)[bl]{highest-speed connections between
 distribution-layer }}

\put(027,32){\makebox(0,0)[bl]{devices
 in different regions}}


\put(47,37){\oval(86,12)} \put(47,37){\oval(85,11)}

\put(44.5,28.5){\makebox(0,0)[bl]{{\bf .~.~.}}}

\put(30,31){\vector(0,-1){04}} \put(30,31){\vector(-1,-1){04}}
\put(30,31){\vector(1,-1){04}}

\put(64,31){\vector(0,-1){04}} \put(64,31){\vector(-1,-1){04}}
\put(64,31){\vector(1,-1){04}}

\put(019,22){\makebox(0,0)[bl]{Distribution  layer
 (work group layer):}}

\put(08,19){\makebox(0,0)[bl]{management of  routing, filtering,
 and QoS policies,}}

\put(02.4,16){\makebox(0,0)[bl]{management of individual
 branch-office WAN connections }}

\put(47,21){\oval(94,12)}

\put(44.5,12.5){\makebox(0,0)[bl]{{\bf .~.~.}}}

\put(30,15){\vector(-1,-1){04}} \put(30,15){\vector(1,-1){04}}
\put(30,15){\vector(0,-1){04}}

\put(64,15){\vector(-1,-1){04}} \put(64,15){\vector(1,-1){04}}
\put(64,15){\vector(0,-1){04}}

\put(02,07){\makebox(0,0)[bl]{Access layer: client nodes,
 workstations, their connections}}

\put(00,06){\line(1,0){94}} \put(00,11){\line(1,0){94}}
\put(00,06){\line(0,1){05}} \put(94,06){\line(0,1){05}}

\end{picture}
\end{center}

 For example,
 the following simplified morphological hierarchical structure
 can be considered (Fig. 8):


 {\bf 0.} System \(S = A\star D \star C\):

 {\bf 1.} Access layer \(A = E \star T\):~
 {\it 1.1.} client nodes \(E\),
 {\it 1.2.} connections \(T\).

 {\bf 2.} Distribution  layer \(D = M \star B \):

 {\it 2.1.} Management \(M = R \star F \star Q\):
 {\it 2.1.1}  routing \(R\),
 {\it 2.1.2.} filtering \(F\),
 {\it 2.1.3.} QoS policies \(Q\).

 {\it 2.2.} Branch-office WAN connections \(B\).

 {\bf 3.} Core layer (core network)
       \(C = H \star K\):

 {\it 3.1.} highest-speed connections between
 distribution-layer devices \(H\),

 {\it 3.2.} core network topology \(K\).


\begin{center}
\begin{picture}(95,41)
\put(001,00){\makebox(0,0)[bl]{Fig. 8.
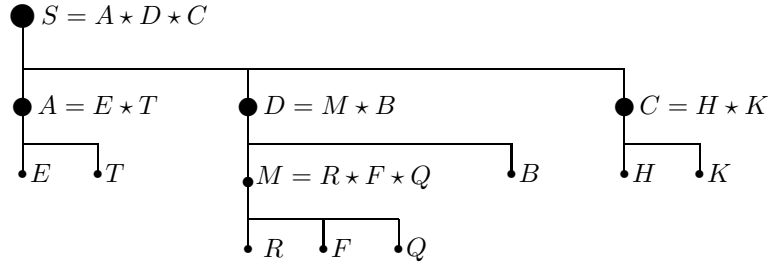
 Structure of three-layer model for enterprise network}}

\put(00,37){\circle*{3.0}}

\put(02.5,36){\makebox(0,0)[bl]{\(S = A\star D \star C\)}}
\put(00,30){\line(0,1){07}} \put(00,30){\line(1,0){80}}

\put(00,25){\circle*{2.5}}

\put(02,24){\makebox(0,0)[bl]{\(A = E \star T\)}}
\put(00,20){\line(0,1){10}} \put(00,20){\line(1,0){10}}

\put(00,16){\line(0,1){04}} \put(00,16){\circle*{1}}
\put(01,15){\makebox(0,0)[bl]{\(E\) }}
\put(10,16){\line(0,1){04}} \put(10,16){\circle*{1}}
\put(11,15){\makebox(0,0)[bl]{\(T\)}}

\put(30,25){\circle*{2.5}}

\put(32,24){\makebox(0,0)[bl]{\(D =M\star B  \) }}

\put(30,20){\line(0,1){10}} \put(30,20){\line(1,0){35}}

\put(65,20){\line(0,-1){04}}

\put(65,16){\circle*{1}} \put(66,15){\makebox(0,0)[bl]{\(B\) }}

\put(30,15){\circle*{1.5}}

\put(31,14){\makebox(0,0)[bl]{\(M =R\star F \star Q \) }}
\put(30,10){\line(0,1){10}} \put(30,10){\line(1,0){20}}

\put(30,06){\line(0,1){04}} \put(30,06){\circle*{1}}
\put(32,05){\makebox(0,0)[bl]{\(R\) }}
\put(40,06){\line(0,1){04}} \put(40,06){\circle*{1}}
\put(41,05){\makebox(0,0)[bl]{\(F\)}}

\put(50,06){\line(0,1){04}} \put(50,06){\circle*{1}}
\put(51,04.5){\makebox(0,0)[bl]{\(Q\) }}

\put(80,25){\circle*{2.5}}

\put(82,24){\makebox(0,0)[bl]{\(C = H \star K\)}}
\put(80,20){\line(0,1){10}} \put(80,20){\line(1,0){10}}

\put(80,16){\line(0,1){04}} \put(80,16){\circle*{1}}
\put(81,15){\makebox(0,0)[bl]{\(H\) }}
\put(90,16){\line(0,1){04}} \put(90,16){\circle*{1}}
\put(91,15){\makebox(0,0)[bl]{\(K\)}}

\end{picture}
\end{center}

 In recent years,
 the following multi-part structure of wireless mobile technologies is examined
 (simplified version)
  \cite{berez02,fag14,gill15,gos15,hoss13,mehta14,sharma13,singh07,yadav17}
  (Fig. 9):

\begin{center}
\begin{picture}(113,89)
\put(012,00){\makebox(0,0)[bl] {Fig. 9.
   Structure of wireless mobile system generations}}

\put(25,84){\circle*{2.7}}

\put(27,83.5){\makebox(0,0)[bl]{\(S =
  P^{1}\star P^{2}\star P^{3}\star P^{4}
 \star P^{5} \star P^{6}\star P^{7} \star P^{8}\)}}

\put(26.5,78){\makebox(0,0)[bl]{\(S^{1G} =
 P^{1}_{1}\star P^{1}_{2}\star P^{3}_{1}\star P^{4}_{1}
 \star P^{5}_{1}\star P^{6}_{1}\star P^{7}_{1}\star P^{8}_{1}\)}}

\put(26.5,73){\makebox(0,0)[bl]{\(S^{2G} =
 P^{1}_{2}\star P^{2}_{2}\star P^{3}_{5}\star P^{4}_{3}
 \star P^{5}_{3}\star P^{6}_{3}\star P^{7}_{1}\star P^{8}_{1} \)}}

\put(26.5,68){\makebox(0,0)[bl]{\(S^{3G} =
 P^{1}_{4}\star P^{2}_{3}\star P^{3}_{6}\star P^{4}_{4}
 \star P^{5}_{4}\star P^{6}_{3}\star P^{7}_{1} \star P^{8}_{1}\)}}

\put(26.5,63){\makebox(0,0)[bl]{\(S^{4G} =
 P^{1}_{5}\star P^{2}_{4}\star P^{3}_{7}\star P^{4}_{4}
 \star P^{5}_{5}\star P^{6}_{6}\star P^{7}_{3}\star P^{8}_{1} \)}}

\put(26.5,58){\makebox(0,0)[bl]{\(S^{5G} =
 P^{1}_{6}\star P^{2}_{5}\star P^{9}_{9}\star P^{4}_{7}
 \star P^{5}_{6}\star P^{6}_{6}\star P^{7}_{3}\star P^{8}_{1} \)}}

\put(26.5,53){\makebox(0,0)[bl]{\(S^{6G} =
 P^{1}_{6}\star P^{2}_{6}\star P^{3}_{9}\star P^{4}_{7}
 \star P^{5}_{8}\star P^{6}_{8}\star P^{7}_{3} \star P^{8}_{1}\)}}

\put(26.5,48){\makebox(0,0)[bl]{\(S^{7G} =
 P^{1}_{6}\star P^{2}_{6}\star P^{3}_{9}\star P^{4}_{7}
 \star P^{5}_{4}\star P^{6}_{8}\star P^{7}_{3} \star P^{8}_{2}\)}}

\put(10,82.5){\makebox(0,0)[bl]{System}}
\put(08.6,79.5){\makebox(0,0)[bl]{structure}}

\put(25,84){\line(0,-1){38}}

\put(02,46){\line(1,0){105}}

\put(02,42){\line(0,1){4}} \put(02,42){\circle*{1.4}}
\put(03.5,42){\makebox(0,0)[bl]{\(P^{1}\)}}

\put(00,37){\makebox(0,0)[bl]{\(P^{1}_{1}\)}}
\put(00,33){\makebox(0,0)[bl]{\(P^{1}_{2}\) }}
\put(00,29){\makebox(0,0)[bl]{\(P^{1}_{3}\)}}
\put(00,25){\makebox(0,0)[bl]{\(P^{1}_{4}\)}}
\put(00,21){\makebox(0,0)[bl]{\(P^{1}_{5}\) }}
\put(00,17){\makebox(0,0)[bl]{\(P^{1}_{6}\)}}

\put(17,42){\line(0,1){4}} \put(17,42){\circle*{1.4}}
\put(18.5,42){\makebox(0,0)[bl]{\(P^{2}\)}}

\put(15,37){\makebox(0,0)[bl]{\(P^{2}_{1}\)}}
\put(15,33){\makebox(0,0)[bl]{\(P^{2}_{2}\)}}
\put(15,29){\makebox(0,0)[bl]{\(P^{2}_{3}\)}}
\put(15,25){\makebox(0,0)[bl]{\(P^{2}_{4}\)}}
\put(15,21){\makebox(0,0)[bl]{\(P^{2}_{5}\)}}
\put(15,17){\makebox(0,0)[bl]{\(P^{2}_{6}\)}}

\put(32,42){\line(0,1){4}} \put(32,42){\circle*{1.4}}
\put(33.5,42){\makebox(0,0)[bl]{\(P^{3}\)}}

\put(30,37){\makebox(0,0)[bl]{\(P^{3}_{1}\)}}
\put(30,33){\makebox(0,0)[bl]{\(P^{3}_{2}\) }}
\put(30,29){\makebox(0,0)[bl]{\(P^{3}_{3}\)}}
\put(30,25){\makebox(0,0)[bl]{\(P^{3}_{4}\)}}
\put(30,21){\makebox(0,0)[bl]{\(P^{3}_{5}\) }}
\put(30,17){\makebox(0,0)[bl]{\(P^{3}_{6}\)}}
\put(30,13){\makebox(0,0)[bl]{\(P^{3}_{7}\)}}
\put(30,09){\makebox(0,0)[bl]{\(P^{3}_{8}\) }}
\put(30,05){\makebox(0,0)[bl]{\(P^{3}_{9}\)}}

\put(47,42){\line(0,1){4}} \put(47,42){\circle*{1.4}}
\put(48.5,42){\makebox(0,0)[bl]{\(P^{4}\)}}

\put(45,37){\makebox(0,0)[bl]{\(P^{4}_{1}\)}}
\put(45,33){\makebox(0,0)[bl]{\(P^{4}_{2}\)}}
\put(45,29){\makebox(0,0)[bl]{\(P^{4}_{3}\)}}
\put(45,25){\makebox(0,0)[bl]{\(P^{4}_{4}\)}}
\put(45,21){\makebox(0,0)[bl]{\(P^{4}_{5}\)}}
\put(45,17){\makebox(0,0)[bl]{\(P^{4}_{6}\)}}
\put(45,13){\makebox(0,0)[bl]{\(P^{4}_{7}\)}}

\put(62,42){\line(0,1){4}} \put(62,42){\circle*{1.4}}
\put(63.5,42){\makebox(0,0)[bl]{\(P^{5}\)}}

\put(60,37){\makebox(0,0)[bl]{\(P^{5}_{1}\)}}
\put(60,33){\makebox(0,0)[bl]{\(P^{5}_{2}\)}}
\put(60,29){\makebox(0,0)[bl]{\(P^{5}_{3}\)}}
\put(60,25){\makebox(0,0)[bl]{\(P^{5}_{4}\)}}

\put(77,42){\line(0,1){4}} \put(77,42){\circle*{1.4}}
\put(78.5,42){\makebox(0,0)[bl]{\(P^{6}\)}}

\put(75,37){\makebox(0,0)[bl]{\(P^{6}_{1}\)}}
\put(75,33){\makebox(0,0)[bl]{\(P^{6}_{2}\)}}
\put(75,29){\makebox(0,0)[bl]{\(P^{6}_{3}\)}}
\put(75,25){\makebox(0,0)[bl]{\(P^{6}_{4}\)}}
\put(75,21){\makebox(0,0)[bl]{\(P^{6}_{5}\)}}
\put(75,17){\makebox(0,0)[bl]{\(P^{6}_{6}\)}}
\put(75,13){\makebox(0,0)[bl]{\(P^{6}_{7}\)}}
\put(75,09){\makebox(0,0)[bl]{\(P^{6}_{8}\)}}

\put(92,42){\line(0,1){4}} \put(92,42){\circle*{1.4}}
\put(93.5,42){\makebox(0,0)[bl]{\(P^{7}\)}}

\put(90,37){\makebox(0,0)[bl]{\(P^{7}_{1}\)}}
\put(90,33){\makebox(0,0)[bl]{\(P^{7}_{2}\)}}
\put(90,29){\makebox(0,0)[bl]{\(P^{7}_{3}\)}}

\put(107,42){\line(0,1){4}} \put(107,42){\circle*{1.4}}
\put(108.5,42){\makebox(0,0)[bl]{\(P^{8}\)}}

\put(105,37){\makebox(0,0)[bl]{\(P^{8}_{1}\)}}
\put(105,33){\makebox(0,0)[bl]{\(P^{8}_{2}\)}}

\end{picture}
\end{center}


 {\bf 0.} Wireless mobile system
   \(S= P^{1} \star P^{2} \star P^{3} \star P^{4} \star P^{5} \star P^{6} \star P^{7} \star P^{8}\):

 {\bf 1.} Part 1 (definition/technology) \(P^{1}\):~
 analog cellular technology \(P^{1}_{1}\), 
 digital cellular technology
 (digital narrow band circuit data)  \(P^{1}_{2}\), 
 packet data \(P^{1}_{3}\), 
 digital broadband packet data \& IP technology \(P^{1}_{4}\), 
 all IP very high throughput \(P^{1}_{5}\), 
  flat IP network \(P^{1}_{6} =P^{1}_{4} \& P^{1}_{5} \). 

 {\bf 2.} Part 2 (data bandwidth/throughput speed/data rates) \(P^{2}\):~
  2 kbps \(P^{2}_{1}\), 
  64 kbps  \(P^{2}_{2}\), 
  400 kbps to 30 Mbps  \(P^{2}_{3}\), 
%
  3-5 Mbps, 100 Mbps (Wifi)  \(P^{2}_{4}\), 
  200 mbps to  1 Gbps  \(P^{2}_{5}\). 
  approx 20 Gbps   \(P^{2}_{6}\). 

 {\bf 3.} Part 3 (service)  \(P^{3}\):~
  mobile telephony (voice) \(P^{3}_{1}\), 
 digital voice  \(P^{3}_{2}\), 
 SMS  \(P^{3}_{3}\), 
 higher capacity packetized data  \(P^{4}_{4}\), 
 digital voice \&   SMS \& higher capacity packetized data
  \(P^{3}_{5}= P^{3}_{2} \&P^{3}_{3} \& P^{3}_{4} \), 
 integrated high quality audio, video and data   \(P^{3}_{6}\), 
 dynamic information access, wearable devices  \(P^{3}_{7}\), 
 AI capability   \(P^{3}_{8}\), 
  dynamic information access, wearable devices \&
 AI capability   \(P^{3}_{9}  = P^{3}_{7} \& P^{3}_{8} \). 

 {\bf 4.} Part 4 (multiplexing/access technology) \(P^{4}\):~
 FDMA  \(P^{4}_{1}\), 
 TDMA  \(P^{4}_{2}\), 
 CDMA  \(P^{4}_{3}\), 
 TDMA \& CDMA  \(P^{4}_{4} = P^{4}_{2} \& P^{4}_{3} \), 
 OFDMA   \(P^{3}_{5}\), 
 LAS-CDMA \(P^{3}_{6}\), 
  OFDMA \& LAS-CDMA  \(P^{3}_{7}= P^{3}_{5}\&P^{3}_{6} \). 

 {\bf 5.} Part 5  (switching) \(P^{5}\):~
 circuit  \(P^{5}_{1}\), 
 packet  \(P^{5}_{2}\), 
 circuit\&packet  \(P^{5}_{3}= P^{5}_{1} \& P^{5}_{2}\), 
 all packet  \(P^{5}_{4}\). 

 {\bf 6.} Part 6 (core network) \(P^{6}\):~
  PSTN \(P^{6}_{1}\), 
  GSM \(P^{6}_{2}\), 
  PSTN \& GSM \(P^{6}_{3}=P^{6}_{1} \& P^{6}_{2}\). 
  packet N/W \(P^{6}_{4}\), 
     Internet \(P^{6}_{5}\), 
  packet N/W \& Internet \(P^{6}_{6}  = P^{6}_{4} \& P^{6}_{5} \), 
  satellite network \(P^{6}_{7}\), 
  packet N/W \& Internet \& satellite network \(P^{6}_{8} = P^{6}_{4}\&P^{6}_{5}\&P^{6}_{7} \). 

 {\bf 7.} Part 7 (handover) \(P^{7}\):~
 horizontal  \(P^{7}_{1}\), 
 vertical  \(P^{7}_{2}\), 
 horizontal \& vertical  \(P^{7}_{3} =P^{7}_{1} \& P^{7}_{2} \). 

 {\bf 8.} Part 8 (satellite functions) \(P^{8}\):~
 none  \(P^{8}_{1}\), 
 satellite roaming  \(P^{8}_{2}\). %

~~

 In this paper,
 the following hierarchical structure
 of wireless mobile technologies is suggested
 (simplified version) (Fig. 10):

\begin{center}
\begin{picture}(150,88)
\put(021,00){\makebox(0,0)[bl] {Fig. 10.
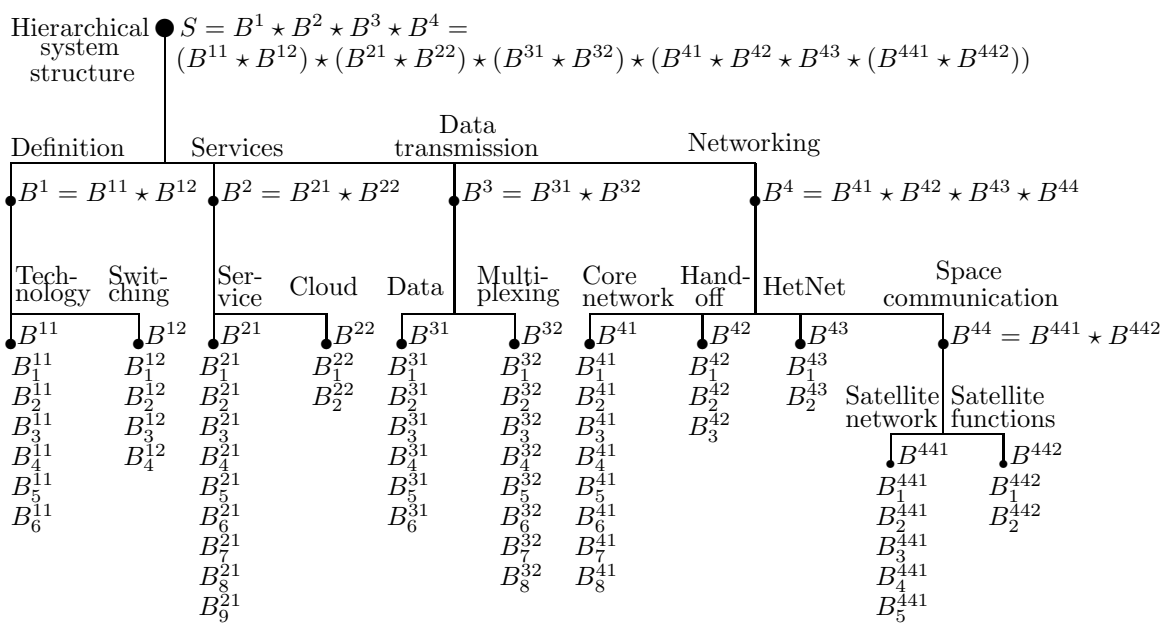
  Hierarchical structure of wireless mobile system generations}}

\put(00,83){\makebox(0,0)[bl]{Hierarchical}}
\put(04,79.7){\makebox(0,0)[bl]{system}}
\put(02.6,77){\makebox(0,0)[bl]{structure}}

\put(20.5,84){\line(0,-1){18}}

\put(20.5,84){\circle*{2.4}}

\put(22.5,83){\makebox(0,0)[bl]{\(S =
  B^{1}\star B^{2}\star B^{3}\star B^{4}= \)}}

\put(22,78){\makebox(0,0)[bl]{\((B^{11}\star B^{12}) \star
(B^{21}\star B^{22}) \star (B^{31}\star B^{32}) \star
  (B^{41}\star B^{42} \star B^{43} \star ( B^{441}\star B^{442} ) )
  \)}}

\put(00,66){\line(1,0){99}}

\put(00,67){\makebox(0,0)[bl]{Definition}}

\put(00,46){\line(0,1){20}} \put(00,61){\circle*{1.4}}
\put(00.9,61){\makebox(0,0)[bl]{\(B^{1}=B^{11}\star B^{12}\)}}

\put(00,46){\line(1,0){17}}

\put(0.6,50){\makebox(0,0)[bl]{Tech-}}
\put(0.6,47){\makebox(0,0)[bl]{nology}}

\put(00,42){\line(0,1){4}} \put(00,42){\circle*{1.4}}
\put(00.8,42){\makebox(0,0)[bl]{\(B^{11}\)}}

\put(00,37){\makebox(0,0)[bl]{\(B^{11}_{1}\)}}
\put(00,33){\makebox(0,0)[bl]{\(B^{11}_{2}\) }}
\put(00,29){\makebox(0,0)[bl]{\(B^{11}_{3}\)}}
\put(00,25){\makebox(0,0)[bl]{\(B^{11}_{4}\)}}
\put(00,21){\makebox(0,0)[bl]{\(B^{11}_{5}\) }}
\put(00,17){\makebox(0,0)[bl]{\(B^{11}_{6}\)}}

\put(13,50){\makebox(0,0)[bl]{Swit-}}
\put(13,47){\makebox(0,0)[bl]{ching}}

\put(17,42){\line(0,1){4}} \put(17,42){\circle*{1.4}}
\put(17.8,42){\makebox(0,0)[bl]{\(B^{12}\)}}

\put(15,37){\makebox(0,0)[bl]{\(B^{12}_{1}\)}}
\put(15,33){\makebox(0,0)[bl]{\(B^{12}_{2}\)}}
\put(15,29){\makebox(0,0)[bl]{\(B^{12}_{3}\)}}
\put(15,25){\makebox(0,0)[bl]{\(B^{12}_{4}\)}}

\put(24,67){\makebox(0,0)[bl]{Services}}

\put(27,46){\line(0,1){20}} \put(27,61){\circle*{1.4}}
\put(27.9,61){\makebox(0,0)[bl]{\(B^{2}=B^{21}\star B^{22}\)}}

\put(27,46){\line(1,0){15}}

\put(27.5,50){\makebox(0,0)[bl]{Ser-}}
\put(27.5,47){\makebox(0,0)[bl]{vice}}

\put(27,42){\line(0,1){4}} \put(27,42){\circle*{1.4}}
\put(27.8,42){\makebox(0,0)[bl]{\(B^{21}\)}}

\put(25,37){\makebox(0,0)[bl]{\(B^{21}_{1}\)}}
\put(25,33){\makebox(0,0)[bl]{\(B^{21}_{2}\) }}
\put(25,29){\makebox(0,0)[bl]{\(B^{21}_{3}\)}}
\put(25,25){\makebox(0,0)[bl]{\(B^{21}_{4}\)}}
\put(25,21){\makebox(0,0)[bl]{\(B^{21}_{5}\) }}
\put(25,17){\makebox(0,0)[bl]{\(B^{21}_{6}\)}}
\put(25,13){\makebox(0,0)[bl]{\(B^{21}_{7}\)}}
\put(25,09){\makebox(0,0)[bl]{\(B^{21}_{8}\) }}
\put(25,05){\makebox(0,0)[bl]{\(B^{21}_{9}\)}}

\put(37,48.5){\makebox(0,0)[bl]{Cloud}}

\put(42,42){\line(0,1){4}} \put(42,42){\circle*{1.4}}
\put(42.8,42){\makebox(0,0)[bl]{\(B^{22}\)}}

\put(40,37){\makebox(0,0)[bl]{\(B^{22}_{1}\)}}
\put(40,33){\makebox(0,0)[bl]{\(B^{22}_{2}\) }}

\put(57,70){\makebox(0,0)[bl]{Data}}
\put(51,67){\makebox(0,0)[bl]{transmission}}

\put(59,46){\line(0,1){20}} \put(59,61){\circle*{1.4}}
\put(59.9,61){\makebox(0,0)[bl]{\(B^{3}=B^{31}\star B^{32}\)}}

\put(52,46){\line(1,0){15}}

\put(50,48.5){\makebox(0,0)[bl]{Data}}

\put(52,42){\line(0,1){4}} \put(52,42){\circle*{1.4}}
\put(52.8,42){\makebox(0,0)[bl]{\(B^{31}\)}}

\put(50,37){\makebox(0,0)[bl]{\(B^{31}_{1}\)}}
\put(50,33){\makebox(0,0)[bl]{\(B^{31}_{2}\)}}
\put(50,29){\makebox(0,0)[bl]{\(B^{31}_{3}\)}}
\put(50,25){\makebox(0,0)[bl]{\(B^{31}_{4}\)}}
\put(50,21){\makebox(0,0)[bl]{\(B^{31}_{5}\)}}
\put(50,17){\makebox(0,0)[bl]{\(B^{31}_{6}\)}}

\put(62,50){\makebox(0,0)[bl]{Multi-}}
\put(62,47){\makebox(0,0)[bl]{plexing}}

\put(67,42){\line(0,1){4}} \put(67,42){\circle*{1.4}}
\put(67.8,42){\makebox(0,0)[bl]{\(B^{32}\)}}

\put(65,37){\makebox(0,0)[bl]{\(B^{32}_{1}\)}}
\put(65,33){\makebox(0,0)[bl]{\(B^{32}_{2}\)}}
\put(65,29){\makebox(0,0)[bl]{\(B^{32}_{3}\)}}
\put(65,25){\makebox(0,0)[bl]{\(B^{32}_{4}\)}}
\put(65,21){\makebox(0,0)[bl]{\(B^{32}_{5}\)}}
\put(65,17){\makebox(0,0)[bl]{\(B^{32}_{6}\)}}
\put(65,13){\makebox(0,0)[bl]{\(B^{32}_{7}\)}}
\put(65,09){\makebox(0,0)[bl]{\(B^{32}_{8}\)}}

\put(90,67){\makebox(0,0)[bl]{Networking}}

\put(99,46){\line(0,1){20}} \put(99,61){\circle*{1.4}}
\put(99.9,61){\makebox(0,0)[bl]{\(B^{4}=B^{41}\star B^{42}\star
  B^{43} \star B^{44}\)}}

\put(77,46){\line(1,0){47}}

\put(76,50){\makebox(0,0)[bl]{Core}}
\put(76,47){\makebox(0,0)[bl]{network}}

\put(77,42){\line(0,1){4}} \put(77,42){\circle*{1.4}}
\put(77.8,42){\makebox(0,0)[bl]{\(B^{41}\)}}

\put(75,37){\makebox(0,0)[bl]{\(B^{41}_{1}\)}}
\put(75,33){\makebox(0,0)[bl]{\(B^{41}_{2}\)}}
\put(75,29){\makebox(0,0)[bl]{\(B^{41}_{3}\)}}
\put(75,25){\makebox(0,0)[bl]{\(B^{41}_{4}\)}}
\put(75,21){\makebox(0,0)[bl]{\(B^{41}_{5}\)}}
\put(75,17){\makebox(0,0)[bl]{\(B^{41}_{6}\)}}
\put(75,13){\makebox(0,0)[bl]{\(B^{41}_{7}\)}}
\put(75,09){\makebox(0,0)[bl]{\(B^{41}_{8}\)}}

\put(89,50){\makebox(0,0)[bl]{Hand-}}
\put(91,47){\makebox(0,0)[bl]{off}}

\put(92,42){\line(0,1){4}} \put(92,42){\circle*{1.4}}
\put(92.8,42){\makebox(0,0)[bl]{\(B^{42}\)}}

\put(90,37){\makebox(0,0)[bl]{\(B^{42}_{1}\)}}
\put(90,33){\makebox(0,0)[bl]{\(B^{42}_{2}\)}}
\put(90,29){\makebox(0,0)[bl]{\(B^{42}_{3}\)}}

\put(100,48.5){\makebox(0,0)[bl]{HetNet}}

\put(105,42){\line(0,1){4}} \put(105,42){\circle*{1.4}}
\put(105.8,42){\makebox(0,0)[bl]{\(B^{43}\)}}

\put(103,37){\makebox(0,0)[bl]{\(B^{43}_{1}\)}}
\put(103,33){\makebox(0,0)[bl]{\(B^{43}_{2}\)}}

\put(123,50){\makebox(0,0)[bl]{Space}}
\put(116,47){\makebox(0,0)[bl]{communication}}

\put(124,42){\line(0,1){4}} \put(124,42){\circle*{1.4}}
\put(124.8,42){\makebox(0,0)[bl]{\(B^{44}=B^{441}\star B^{442}\)
 }}
\put(124,42){\line(0,-1){12}}

\put(111,34){\makebox(0,0)[bl]{Satellite}}
\put(111,31){\makebox(0,0)[bl]{network}}

\put(117,26){\line(0,1){4}} \put(117,26){\circle*{1.2}}
\put(117.8,26){\makebox(0,0)[bl]{\(B^{441}\)}}

\put(115,21){\makebox(0,0)[bl]{\(B^{441}_{1}\)}}
\put(115,17){\makebox(0,0)[bl]{\(B^{441}_{2}\)}}
\put(115,13){\makebox(0,0)[bl]{\(B^{441}_{3}\)}}
\put(115,09){\makebox(0,0)[bl]{\(B^{441}_{4}\)}}
\put(115,05){\makebox(0,0)[bl]{\(B^{441}_{5}\)}}

\put(117,30){\line(1,0){15}}

\put(125,34){\makebox(0,0)[bl]{Satellite}}
\put(125,31){\makebox(0,0)[bl]{functions}}

\put(132,30){\line(0,-1){4}} \put(132,26){\circle*{1.2}}
\put(132.8,26){\makebox(0,0)[bl]{\(B^{442}\)}}

\put(130,21){\makebox(0,0)[bl]{\(B^{442}_{1}\)}}
\put(130,17){\makebox(0,0)[bl]{\(B^{442}_{2}\)}}

\end{picture}
\end{center}


 {\bf 0.} Wireless mobile system
 \(S= B^{1} \star B^{2} \star B^{3} \star B^{4}\):

 {\bf 1.} Definition \(  B^{1} = B^{11} \star B^{12}\):~

 {\bf 1.1.} technology \(B^{11}\):~
 analog cellular technology \(B^{11}_{1}\), 
 digital cellular technology
 (digital narrow band circuit data)  \(B^{11}_{2}\), 
 packet data \(B^{11}_{3}\), 
 digital broadband packet data \& IP technology \(B^{11}_{4}\), 
 all IP very high throughput \(B^{11}_{5}\), 
  flat IP network \(B^{11}_{6} =B^{11}_{4} \& B^{11}_{5} \). 

 {\bf 1.2.} switching \(B^{12}\):~
 circuit  \(B^{12}_{1}\), 
 packet  \(B^{12}_{2}\), 
 circuit\&packet  \(B^{12}_{3}= B^{12}_{1} \& B^{12}_{2}\), 
 all packet  \(B^{12}_{4}\). 

 {\bf 2.} Services \(  B^{2} = B^{21} \star B^{22}\):~

 {\bf 2.1.} service  \(B^{21}\):~
 mobile telephony (voice) \(B^{21}_{1}\), 
 digital voice  \(B^{21}_{2}\), 
 SMS  \(B^{21}_{3}\), 
 higher capacity packetized data  \(B^{21}_{4}\), 
 digital voice \&   SMS \& higher capacity packetized data
  \(B^{21}_{5}= B^{21}_{2} \&B^{21}_{3} \& B^{21}_{4} \), 
 integrated high quality audio, video and data   \(B^{21}_{6}\), 
 dynamic information access, wearable devices  \(B^{21}_{7}\), 
 AI capability   \(B^{21}_{8}\), 
  dynamic information access, wearable devices \&
 AI capability   \(B^{21}_{9}  = B^{21}_{7} \& B^{21}_{8} \). 

 {\bf 2.2.} cloud computing  \(B^{22}\):~
  none \(B^{22}_{1}\), 
  cloud computing  \(B^{22}_{2}\), 

 {\bf 3.} Data transmission \& access
    \(  B^{3} = B^{31} \star B^{32}\):~

 {\bf 3.1.} data bandwidth/throughput speed/data rates \(B^{31}\):~
  2 kbps \(B^{31}_{1}\), 
  64 kbps  \(B^{31}_{2}\), 
  400 kbps to 30 Mbps  \(B^{31}_{3}\), 
%
  3-5 Mbps, 100 Mbps (Wifi)  \(B^{31}_{4}\), 
  200 mbps to  1 Gbps  \(B^{31}_{5}\), 
  approx 20 Gbps   \(B^{31}_{6}\). 

 {\bf 3.2.} multiplexing/access technology \(B^{32}\):~
 FDMA  \(B^{32}_{1}\), 
 TDMA  \(B^{32}_{2}\), 
 FDMA \& TDMA  \(B^{32}_{3} =B^{32}_{1} \& B^{32}_{2} \), 
 CDMA  \(B^{32}_{4}\), 
 TDMA \& CDMA  \(B^{32}_{5} = B^{32}_{2} \& B^{32}_{4} \), 
 OFDMA   \(B^{32}_{6}\), 
 LAS-CDMA \(B^{32}_{7}\), 
  OFDMA \& LAS-CDMA  \(B^{32}_{8}= B^{32}_{6}\&B^{32}_{7} \). 

 {\bf 4.} Networking
    \(  B^{4} = B^{41} \star B^{42} \star B^{43} \star  B^{44}\):~

 {\bf 4.1.} core network \(B^{41}\):~
  PSTN \(B^{41}_{1}\), 
  GSM \(B^{41}_{2}\), 
  PSTN \& GSM \(B^{41}_{3}=B^{41}_{1} \& B^{41}_{2}\). 
  packet N/W \(B^{41}_{4}\), 
     Internet \(B^{41}_{5}\), 
  packet N/W \& Internet \(B^{41}_{6}  = B^{41}_{4} \& B^{41}_{5} \), 
  satellite network \(B^{41}_{7}\), 
  packet N/W \& Internet \& satellite network
    \(B^{41}_{8} = B^{41}_{4}\&B^{41}_{5}\&B^{41}_{7} \). 

 {\bf 4.2.} handoff \(B^{42}\):~
 horizontal  \(B^{42}_{1}\), 
 vertical  \(B^{42}_{2}\), 
 horizontal \& vertical  \(B^{42}_{3} =B^{42}_{1} \& B^{42}_{2} \). 

 {\bf 4.3.} heterogeneous networks (HetNets) \(B^{43}\):~
 none  \(B^{43}_{1}\), 
 aggregation of different networks (HetNets)  \(B^{43}_{2}\). 

 {\bf 4.4.} space communication \(B^{44} = B^{441} \& B^{442}\):~

 {\it 4.4.1.} satellite network \(B^{441}\):~
 none \(B^{441}_{1}\), 
 telecommunication network \(B^{441}_{2}\),
 earth imaging \(B^{441}_{3}\),
 navigation \(B^{441}_{4}\),
 telecommunication \& earth imaging \& navigation
  \(B^{441}_{5} = B^{441}_{2} \& B^{441}_{3} \& B^{441}_{4}\). 

 {\it 4.4.2.} satellite functions \(B^{442}\):~
 none  \(B^{442}_{1}\), 
 satellite roaming  \(B^{442}_{2}\). 

~~

 As a result,
 the following modular system descriptions can be examined:

 (1) \(S^{1G} =
 (B^{11}_{1} \star B^{12}_{1}) \star (B^{21}_{1} \star B^{22}_{1})
 \star (B^{31}_{1} \star B^{32}_{3}) \star
  (B^{41}_{1}\star B^{42}_{1} \star B^{43}_{1}
  \star ( B^{441}_{1}\star B^{442}_{1} ) ) \),

 (2) \(S^{2G} =
 (B^{11}_{2} \star B^{12}_{3}) \star (B^{21}_{5} \star B^{22}_{1})
 \star (B^{31}_{2} \star B^{32}_{4}) \star
  (B^{41}_{3}\star B^{42}_{1} \star B^{43}_{1}
  \star ( B^{441}_{1}\star B^{442}_{1} ) ) \),

 (3) \(S^{3G} =
 (B^{11}_{4} \star B^{12}_{2}) \star (B^{21}_{6} \star B^{22}_{1})
 \star (B^{31}_{3} \star B^{32}_{5}) \star
  (B^{41}_{4}\star B^{42}_{1} \star B^{43}_{1}
  \star ( B^{441}_{1}\star B^{442}_{1} ) )\),

 (4) \(S^{4G} =
 (B^{11}_{5} \star B^{12}_{4}) \star (B^{21}_{7} \star B^{22}_{1})
 \star (B^{31}_{4} \star B^{32}_{5}) \star
  (B^{41}_{5}\star B^{42}_{3} \star B^{43}_{1}
  \star ( B^{441}_{1}\star B^{442}_{1} ) ) \),

 (5) \(S^{5G} =
 (B^{11}_{6} \star B^{12}_{4}) \star (B^{21}_{8} \star B^{22}_{2})
 \star (B^{31}_{5} \star B^{32}_{5}) \star
  (B^{41}_{5}\star B^{42}_{3} \star B^{43}_{2}
  \star ( B^{441}_{1}\star B^{442}_{1} ) )\),

 (6) \(S^{6G} =
 (B^{11}_{6} \star B^{12}_{4}) \star (B^{21}_{9} \star B^{22}_{2})
 \star (B^{31}_{6} \star B^{32}_{8}) \star
  (B^{41}_{8}\star B^{42}_{3} \star B^{43}_{2}
  \star ( B^{441}_{5}\star B^{442}_{1} ) ) \),

 (7) \(S^{7G} =
 (B^{11}_{6} \star B^{12}_{4}) \star (B^{21}_{9} \star B^{22}_{2})
 \star (B^{31}_{6} \star B^{32}_{8}) \star
  (B^{41}_{8}\star B^{42}_{3} \star B^{43}_{2}
  \star ( B^{441}_{5}\star B^{442}_{2} ) ) \).

\section{Illustrative example of two-stage system improvement}

 In this section,
 an illustrative example of two-stage system improvement
 for 5G communication technology is described:~
 \(S^{5G} \Longrightarrow (S^{5G^{adv^{1}}} \rightarrow S^{5G^{adv^{2}}})\),
 where \(S^{5G^{adv^{1}}}\)  is the  system after the improvement stage 1,
 \(S^{5G^{adv^{2}}}\)  is the  system after the improvement stage 2.
 This kind of the improvement problem may be of interest to a
 communication company/organization.

 Note, seven basic combinatorial engineering frameworks
 (design of system hierarchical model,
 system evaluation,
 detection of system bottlenecks,
  system design,
  system improvement,
  multistage system design,
  combinatorial modeling of system evolution,
  system forecasting)
  for
 modular systems have been suggested in \cite{lev13intro,lev14plat,lev15}.
 The considered two stage scheme (framework)
 for two-stage modular system improvement (or forecasting) is
 depicted in Fig. 11.

\begin{center}
\begin{picture}(135,75)
\put(20,00){\makebox(0,0)[bl]{Fig. 11.
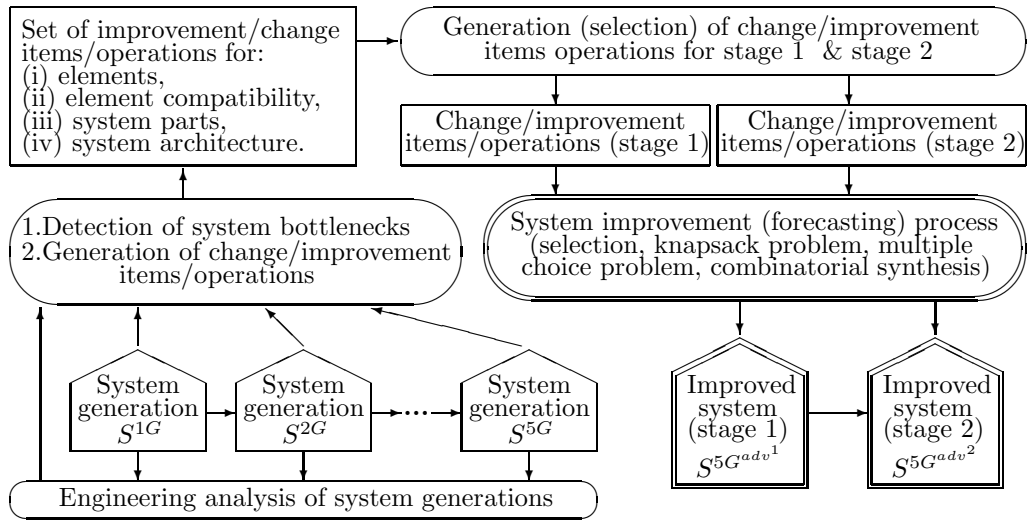
 Two-stage system improvement (forecasting) scheme}}

\put(00,52){\line(1,0){46}} \put(00,73){\line(1,0){46}}
\put(00,52){\line(0,1){21}} \put(46,52){\line(0,1){21}}

\put(01.5,68){\makebox(0,0)[bl]{Set of improvement/change}}
\put(01.5,65){\makebox(0,0)[bl]{items/operations for:}}
\put(01.5,62){\makebox(0,0)[bl]{(i) elements,}}
\put(01.5,59){\makebox(0,0)[bl]{(ii) element compatibility,}}
\put(01.5,56){\makebox(0,0)[bl]{(iii) system parts,}}
\put(01.5,53){\makebox(0,0)[bl]{(iv) system architecture.}}

\put(23,47.5){\vector(0,1){4}} \put(46,68.5){\vector(1,0){05.5}}
\put(30,40.5){\oval(60,14)}

\put(01.5,42){\makebox(0,0)[bl]{1.Detection of system
 bottlenecks}}

\put(01.5,38.5){\makebox(0,0)[bl]{2.Generation of
 change/improvement}}

\put(14.5,35.5){\makebox(0,0)[bl]{items/operations}}

\put(04,10){\vector(0,1){23}} \put(39,28){\vector(-1,1){5}}
\put(68,28){\vector(-4,1){20}} \put(17,28){\vector(0,1){5}}

\put(39,07.5){\oval(78,05)}

\put(06.5,06){\makebox(0,0)[bl]{Engineering analysis of system
 generations}}

\put(17,14){\vector(0,-1){4}} \put(39,14){\vector(0,-1){4}}
\put(69,14){\vector(0,-1){4}}


\put(08,14){\line(1,0){18}}

\put(08,14){\line(0,1){09}} \put(26,14){\line(0,1){09}}
\put(08,23){\line(2,1){09}} \put(26,23){\line(-2,1){09}}

\put(11.5,21){\makebox(0,0)[bl]{System }}
\put(09,18){\makebox(0,0)[bl]{generation}}
\put(13,15){\makebox(0,0)[bl]{ \(S^{1G}\)}}

\put(26,19){\vector(1,0){4}}


\put(30,14){\line(1,0){18}}

\put(30,14){\line(0,1){09}} \put(48,14){\line(0,1){09}}
\put(30,23){\line(2,1){09}} \put(48,23){\line(-2,1){09}}

\put(33.5,21){\makebox(0,0)[bl]{System }}
\put(31,18){\makebox(0,0)[bl]{generation}}
\put(35,15){\makebox(0,0)[bl]{ \(S^{2G}\)}}

\put(48,19){\vector(1,0){4}}


\put(52.3,18.8){\makebox(0,0)[bl]{{\bf ...}}}

\put(56,19){\vector(1,0){4}}


\put(60,14){\line(1,0){18}}

\put(60,14){\line(0,1){09}} \put(78,14){\line(0,1){09}}
\put(60,23){\line(2,1){09}} \put(78,23){\line(-2,1){09}}

\put(63.5,21){\makebox(0,0)[bl]{System }}
\put(61,18){\makebox(0,0)[bl]{generation}}
\put(65,15){\makebox(0,0)[bl]{ \(S^{5G}\)}}

\put(93.5,68.5){\oval(83,9)}

\put(57.5,68.5){\makebox(0,0)[bl]{Generation (selection) of
 change/improvement}}
\put(63,65.5){\makebox(0,0)[bl]{items operations for stage 1 ~\&
 stage 2}}

\put(72.5,64){\vector(0,-1){4}} \put(111.5,64){\vector(0,-1){4}}

\put(52,52){\line(1,0){41}} \put(52,60){\line(1,0){41}}
\put(52,52){\line(0,1){08}} \put(93,52){\line(0,1){08}}

\put(56.5,56){\makebox(0,0)[bl]{Change/improvement}}
\put(52.5,53){\makebox(0,0)[bl]{items/operations (stage 1)}}

\put(72.5,52){\vector(0,-1){4}}

\put(94,52){\line(1,0){41}} \put(94,60){\line(1,0){41}}
\put(94,52){\line(0,1){08}} \put(135,52){\line(0,1){08}}

\put(98,56){\makebox(0,0)[bl]{Change/improvement}}
\put(94.5,53){\makebox(0,0)[bl]{items/operations (stage 2)}}

\put(111.5,52){\vector(0,-1){4}}

\put(99,41){\oval(71,13)} \put(99,41){\oval(72,14)}

\put(66.5,42.5){\makebox(0,0)[bl]{System improvement (forecasting)
  process}}

 \put(69,39.5){\makebox(0,0)[bl]{(selection, knapsack problem, multiple}}

\put(68,36.5){\makebox(0,0)[bl]{choice problem, combinatorial
 synthesis)}}

\put(97,34){\vector(0,-1){4.5}} \put(123,34){\vector(0,-1){4.5}}


\put(88,09){\line(1,0){18}}

\put(88,09){\line(0,1){15.5}} \put(106,09){\line(0,1){15.5}}
\put(88,24.5){\line(2,1){09}} \put(106,24.5){\line(-2,1){09}}

\put(88.5,09.5){\line(1,0){17}}

\put(88.5,09.5){\line(0,1){014.5}}
\put(105.5,09.5){\line(0,1){014.5}}

\put(88.5,24){\line(2,1){08.5}} \put(105.5,24){\line(-2,1){08.5}}

\put(90,21){\makebox(0,0)[bl]{Improved }}
\put(91.5,18){\makebox(0,0)[bl]{system}}
\put(90,15){\makebox(0,0)[bl]{(stage \(1\))}}
\put(90,10.5){\makebox(0,0)[bl]{ \(S^{5G^{adv^{1}}}\)}}

\put(106,19){\vector(1,0){8}}


\put(114,09){\line(1,0){18}}

\put(114,09){\line(0,1){15.5}} \put(132,09){\line(0,1){15.5}}
\put(114,24.5){\line(2,1){09}} \put(132,24.5){\line(-2,1){09}}

\put(114.5,09.5){\line(1,0){17}}

\put(114.5,09.5){\line(0,1){14.5}}
\put(131.5,09.5){\line(0,1){14.5}}

\put(114.5,24){\line(2,1){08.5}} \put(131.5,24){\line(-2,1){08.5}}

\put(116,21){\makebox(0,0)[bl]{Improved}}
\put(117.5,18){\makebox(0,0)[bl]{system}}
\put(116,15){\makebox(0,0)[bl]{(stage \(2\))}}
\put(116,10.5){\makebox(0,0)[bl]{ \(S^{5G^{adv^{2}}}\)}}

\end{picture}
\end{center}

 In general, it is possible to generate (e.g., engineering
 analysis) a set of change/improvement activities (operations), for example as in
 Table 7:~ \(\overline{O} = \{ O_{1},...,O_{\iota},...,O_{17} \}\).

\newpage
\begin{center}
 {\bf Table 7.} Generated improvement
  activities  (based on data from Table 3, Table 4) \\
\begin{tabular}{| c |  l | l |}
\hline

 No.& Change/improvement (forecasting) activity & Notation \\

\hline

 1.&Implementation of  central architecture:&\\
 1.1 &(a) cloud radio-access networks (RAN) based on SDR&\(O_{1}\)\\

   &  and coordinated central controllers&\\

 1.2  & (b) cloud basic networks CN based on SDN&\(O_{2}\)\\

 2.&Multidimensional antennas MIMO&\(O_{3}\)\\

 3.&Flexibility, adaptivity, heterogeneity:&\\

 3.1.&(i) flexible common usage of frequency resources &\(O_{4}\)\\

 3.2.&(ii) terminal and network heterogeneity (different types of access
 networks,
 &\(O_{5}\)\\
    & e.g., WiMAX, WiFi, UMTS)&  \\

 3.3.&(iii) allocation and management of resources in heterogeneous networks
 & \(O_{6}\)\\

 3.4.&(iv)  inter-network joint work for different radio-access
 technologies& \(O_{7}\) \\

 3.5.&(v)  self-adaptation and self-optimization networks &\(O_{8}\)\\

 3.6.&(vi) smart homes, smart cities, smart villages&\(O_{9}\)\\

 4.& Device-centric architectures& \(O_{10}\)\\

 5.& Very wide area coverage &  \(O_{11}\)\\

 6.&User personalization (high data transfer rates, access
    to large  repository & \(O_{12}\) \\

  & of data and services, flexibility)&\\

 7.& Interoperability (unified global standard, global mobility and service
     & \(O_{13}\) \\

  &portability,
  i.e., different services from different service providers)&\\

 8.&Network convergence (convergence with both devices and services)
     &\(O_{14}\)\\

 9.& Lower power consumption&\(O_{15}\)  \\

 10.&Ultra fast access of Internet&\(O_{16}\)\\

 11.&Satellite to satellite communication&\(O_{17}\)\\

\hline
\end{tabular}
\end{center}

 Here, a modular system improvement is examined.
 The initial technology generation under examination is:~
%
  \(S^{5G} =
 (B^{11}_{6} \star B^{12}_{4}) \star (B^{21}_{8} \star B^{22}_{2})
 \star (B^{31}_{5} \star B^{32}_{5}) \star
  (B^{41}_{5}\star B^{42}_{3} \star B^{43}_{2}
  \star ( B^{441}_{1}\star B^{442}_{1} ) )\).
%
 The following technology components of \(S^{5G}\) can be are considered
 for the change/improvement (or forecasting):
 (1) service~ \(B^{21}_{8}\);
 (2) data bandwidth/througput speed/data rates~  \(B^{31}_{5}\);
 (3) multiplexing/access technology~ \(B^{32}_{5}\);
 (4) core network~  \(B^{41}_{5}\);
 (5) satellite network~  \(B^{441}_{1}\); and
 (6) satellite functions~ \(B^{442}_{1}\).
 Thus, two series system improvement (forecasting) problems
 can be considered (a simplified case, knapsack-like models as multiple choice
 problems): for stage 1 and for stage 2.

~~

 {\bf Problem 1.}
 Structure of composite improvement  \(S^{5G} \Rightarrow  S^{5G^{adv^{1}}}\)  for stage 1
  is depicted in Fig. 12.
 Descriptions of the improvement operations
 and their estimates are contained in Table 8.

 The corresponding multiple choice model for
  \(S^{5G} \Rightarrow  S^{5G^{adv^{1}}}\)
 is
 (\(q_{1}=2\), \(q_{2}=2\), \(q_{3}=4\),
 \(q_{4}=4\),  \(q_{5}=5\),
  \(b_{1}^{constr} = 19.0\)):
 \[
 \max \sum_{\iota=1}^{6} \sum_{j=1}^{q_{\iota}} x_{\iota,j} c_{\iota,j} ~~~~
 s.t. ~~~\sum_{\iota=1}^{6}  \sum_{j=1}^{q_{\iota}}  x_{\iota,j} b_{\iota,j}
 \leq  b_{1}^{constr},
 ~~ \sum_{j=1}^{q_{\iota}} x_{\iota,j} \leq 1 ~~
 \forall \iota=\overline{1,6}, ~ \forall j=\overline{1,q_{\iota}}
 ~~~~ \forall x_{\iota,j} \in \{0,1\}.\]
 A simplified greedy heuristic is used
 (i.e., series packing of items via value
 \( \frac{ c_{\iota,j}} { b_{\iota,j} } \).)

\begin{center}
\begin{picture}(95,44)
\put(04.5,00){\makebox(0,0)[bl] {Fig. 12.
   Composite improvement  \(S^{5G} \Rightarrow  S^{5G^{adv^{1}}}\)}}

\put(25,39){\circle*{2.7}}

\put(27.4,38.5){\makebox(0,0)[bl]{\(I^{1} =
  U^{1}\star U^{2}\star U^{3}\star U^{4} \star U^{5}\)}}

\put(26.5,33){\makebox(0,0)[bl]{\(I^{1}_{1} =
  U^{1}_{2} \star U^{2}_{2} \star U^{3}_{3} \star U^{4}_{3} \star U^{5}_{2}
 \)}}

\put(04,37.5){\makebox(0,0)[bl]{Composite}}
\put(02,34.5){\makebox(0,0)[bl]{improvement }}
\put(05,31.5){\makebox(0,0)[bl]{(stage 1)}}

\put(25,39){\line(0,-1){08}}

\put(02,31){\line(1,0){80}}

\put(02,27){\line(0,1){4}} \put(02,27){\circle*{1.4}}
\put(03.5,27){\makebox(0,0)[bl]{\(U^{1}\)}}

\put(00,22){\makebox(0,0)[bl]{\(U^{1}_{1}\)}}
\put(00,18){\makebox(0,0)[bl]{\(U^{1}_{2}\) }}

\put(22,27){\line(0,1){4}} \put(22,27){\circle*{1.4}}
\put(23.5,27){\makebox(0,0)[bl]{\(U^{2}\)}}

\put(20,22){\makebox(0,0)[bl]{\(U^{2}_{1}\)}}
\put(20,18){\makebox(0,0)[bl]{\(U^{2}_{2}\)}}

\put(42,27){\line(0,1){4}} \put(42,27){\circle*{1.4}}
\put(43.5,27){\makebox(0,0)[bl]{\(U^{3}\)}}

\put(40,22){\makebox(0,0)[bl]{\(U^{3}_{1}\)}}
\put(40,18){\makebox(0,0)[bl]{\(U^{3}_{2}\) }}
\put(40,14){\makebox(0,0)[bl]{\(U^{3}_{3}\)}}
\put(40,10){\makebox(0,0)[bl]{\(U^{3}_{4}\)}}

\put(62,27){\line(0,1){4}} \put(62,27){\circle*{1.4}}
\put(63.5,27){\makebox(0,0)[bl]{\(U^{4}\)}}

\put(60,22){\makebox(0,0)[bl]{\(U^{4}_{1}\)}}
\put(60,18){\makebox(0,0)[bl]{\(U^{4}_{2}\)}}
\put(60,14){\makebox(0,0)[bl]{\(U^{4}_{3}\)}}
\put(60,10){\makebox(0,0)[bl]{\(U^{4}_{4}\)}}

\put(82,27){\line(0,1){4}} \put(82,27){\circle*{1.4}}
\put(83.5,27){\makebox(0,0)[bl]{\(U^{5}\)}}

\put(80,22){\makebox(0,0)[bl]{\(U^{5}_{1}\)}}
\put(80,18){\makebox(0,0)[bl]{\(U^{5}_{2}\)}}
\put(80,14){\makebox(0,0)[bl]{\(U^{5}_{3}\)}}
\put(80,10){\makebox(0,0)[bl]{\(U^{5}_{4}\)}}
\put(80,06){\makebox(0,0)[bl]{\(U^{5}_{5}\)}}

\end{picture}
\end{center}

\newpage
\begin{center}
 {\bf Table 8.} Change operations for \(S^{5G} \Rightarrow  S^{5G^{adv^{1}}}\) (stage 1)   \\
\begin{tabular}{|c|l|c|l|c|c |c| }
\hline
 No.&System component&Alternative &Change/improvement&Binary&Profit&Cost\\
 \(\iota\)& &under change&item/operation&variable&\(c_{\iota,j}\)&\(b_{\iota,j}\)\\
\hline

 1.&\(B^{21}\) (service)
  &\(B^{21}_{8}\)&\(U^{1}_{1}:\) None &\(x_{1,1}\)&\(0.0\)&\(0.0\)\\

  &&\(B^{21}_{8}\)&\(U^{1}_{2}: B^{21}_{8} \rightarrow \ B^{21}_{9}\)&\(x_{1,2}\)&\(2.0\)&\(3.0\)\\

\hline
 2.& \(B^{31}\) (data bandwidth/
  &\(B^{31}_{5}\)&\(U^{2}_{1}:\) None &\(x_{2,1}\)&\(0.0\)&\(0.0\)\\

  &througput speed/&\(B^{31}_{5}\)&\(U^{2}_{2}:
  B^{31}_{5} \rightarrow B^{31}_{6}\)&\(x_{2,2}\)&\(4.0\)&\(5.0\)\\

  &data rates)  &&&&&\\

\hline
 3.& \(B^{32}\) (multiplexing/
  &\(B^{32}_{5}\)&\(U^{3}_{1}:\) None &\(x_{3,1}\)&\(0.0\)&\(0.0\)\\

  &access technology)&\(B^{32}_{5}\)&\(U^{3}_{2}:
   B^{32}_{5} \rightarrow B^{32}_{6}\)&\(x_{3,2}\)&\(1.0\)&\(2.0\)\\ 

  &                 &\(B^{32}_{5}\)&\(U^{3}_{3}:
  B^{32}_{5} \rightarrow B^{32}_{7}\)&\(x_{3,3}\)&\(3.6\)&\(4.0\)\\ 

  &                 &\(B^{32}_{5}\)&\(U^{3}_{4}:
  B^{32}_{5} \rightarrow B^{32}_{8}\)&\(x_{3,4}\)&\(3.6\)&\(6.0\)\\ 

\hline
 4.& \(B^{41}\) (core network)
  &\(B^{41}_{5}\)&\(U^{4}_{1}:\) None &\(x_{4,1}\)&\(0.0\)&\(0.0\)\\

  & &\(B^{41}_{5}\)&\(U^{4}_{2}: B^{41}_{5} \rightarrow
  B^{41}_{6}\)&\(x_{4,2}\)&\(3.6\)&\(6.0\)\\ 

  & &\(B^{41}_{5}\)&\(U^{4}_{3}:
  B^{41}_{5} \rightarrow B^{41}_{7}\)&\(x_{4,3}\)&\(7.0\)&\(7.0\)\\ 

  & &\(B^{41}_{5}\)&\(U^{4}_{4}:
  B^{41}_{5} \rightarrow B^{41}_{8}\)&\(x_{4,4}\)&\(9.0\)&\(12.0\)\\ 

\hline
 5.& \(B^{441}\) (satellite  network)
  &\(B^{441}_{1}\)&\(U^{5}_{1}:\) None &\(x_{5,1}\)&\(0.0\)&\(0.0\)\\

  & &\(B^{441}_{1}\)&\(U^{5}_{2}:
  B^{441}_{1} \rightarrow B^{441}_{2}\)&\(x_{5,2}\)&\(5.0\)&\(5.0\)\\ 

  & &\(B^{441}_{1}\)&\(U^{5}_{3}:
  B^{441}_{1} \rightarrow B^{441}_{3}\)&\(x_{5,3}\)&\(5.6\)&\(7.0\)\\ 

  & &\(B^{441}_{1}\)&\(U^{5}_{4}:
  B^{441}_{1} \rightarrow B^{441}_{4}\)&\(x_{5,4}\)&\(6.0\)&\(8.0\)\\ 

  & &\(B^{441}_{1}\)&\(U^{5}_{5}:
  B^{441}_{1} \rightarrow B^{441}_{5}\)&\(x_{5,5}\)&\(14.0\)&\(20.0\)\\

\hline
\end{tabular}
\end{center}

%
%
 The resultant solution is
 (\( I^{1}_{1}=
 U^{1}_{2} \star U^{2}_{2} \star U^{3}_{3} \star U^{4}_{3} \star U^{6}_{2}\)):~

%
  \(S^{5G^{adv^{1}}} =
 (B^{11}_{6} \star B^{12}_{4}) \star (B^{21}_{9} \star B^{22}_{2})
 \star (B^{31}_{6} \star B^{32}_{7}) \star
  (B^{41}_{7}\star B^{42}_{3} \star B^{43}_{2}
  \star ( B^{441}_{2}\star B^{442}_{1} ) )\).

 This modular system is a basis for the improvement (forecasting) at stage 2.

~~

 {\bf Problem 2.}
 Structure of composite improvement
 \(S^{5G^{adv^{1}}} \Rightarrow  S^{5G^{adv^{2}}}\)  for stage 2
  is depicted in Fig. 13.
 Descriptions of the improvement operations
 and their estimates are contained in Table 9.
 The corresponding multiple choice model for
  \( S^{5G^{adv^{1}}} \Rightarrow  S^{5G^{adv^{2}}}\)
 is
 (\(q_{1}=2\), \(q_{2}=2\), \(q_{3}=4\),
 \(q_{4}=2\),
  \(b_{2}^{constr}=17.5\)):
 \[
 \max \sum_{\iota=1}^{4} \sum_{j=1}^{q_{\iota}} y_{\iota,j} c_{\iota,j} ~~~~
 s.t. ~~~\sum_{\iota=1}^{4}  \sum_{j=1}^{q_{\iota}}  y_{\iota,j} b_{\iota,j} < b_{2}^{constr},
 ~~ \sum_{j=1}^{q_{\iota}} y_{\iota,j} \leq 1 ~~
 \forall \iota=\overline{1,4}, ~ \forall j=\overline{1,q_{\iota}}
 ~~~~ \forall y_{\iota,j} \in \{0,1\}.\]
%
%
 Here, a simplified greedy heuristic is used
 (i.e., series packing of items via value
 \(  \frac{ c_{\iota,j} } { b_{\iota,j}} \)).
 The resultant solution is:~
 (\( I^{2}_{1}=
 V^{1}_{2} \star V^{2}_{2} \star V^{3}_{2} \star V^{4}_{1} \)):~

%
  \(S^{5G^{adv^{2}}} =
 (B^{11}_{6} \star B^{12}_{4}) \star (B^{21}_{9} \star B^{22}_{2})
 \star (B^{31}_{6} \star B^{32}_{8}) \star
  (B^{41}_{8}\star B^{42}_{3} \star B^{43}_{2}
  \star ( B^{441}_{3}\star B^{442}_{1} ) )\).

 Thus,
 the following two-stage improvement (forecasting) strategy
 (as a chain) is obtained):

 \(\Upsilon:~~
  < S^{5G^{adv}} \Longrightarrow S^{5G^{adv^{1}}} \Rightarrow
 S^{5G^{adv^{2}}} > \).

\begin{center}
\begin{picture}(90,39)
\put(00,00){\makebox(0,0)[bl] {Fig. 13.
   Composite improvement  \( S^{5G^{adv^{1}}} \Rightarrow  S^{5G^{adv^{2}}}\)}}

\put(25,34){\circle*{2.7}}

\put(27.4,33.5){\makebox(0,0)[bl]{\(I^{2} =
  V^{1}\star V^{2}\star V^{3}\star V^{4}\)}}

\put(26.5,28){\makebox(0,0)[bl]{\(I^{2}_{1} =
 V^{1}_{1}\star V^{1}_{2}\star V^{3}_{1}\star V^{4}_{1}\)}}

\put(04,32.5){\makebox(0,0)[bl]{Composite}}
\put(02,29.5){\makebox(0,0)[bl]{improvement }}
\put(05,26.5){\makebox(0,0)[bl]{(stage 2)}}

\put(25,34){\line(0,-1){08}}

\put(02,26){\line(1,0){75}}

\put(02,22){\line(0,1){4}} \put(02,22){\circle*{1.4}}
\put(03.5,22){\makebox(0,0)[bl]{\(V^{1}\)}}

\put(00,17){\makebox(0,0)[bl]{\(V^{1}_{1}\)}}
\put(00,13){\makebox(0,0)[bl]{\(V^{1}_{2}\) }}

\put(27,22){\line(0,1){4}} \put(27,22){\circle*{1.4}}
\put(28.5,22){\makebox(0,0)[bl]{\(V^{2}\)}}

\put(25,17){\makebox(0,0)[bl]{\(V^{2}_{1}\)}}
\put(25,13){\makebox(0,0)[bl]{\(V^{2}_{2}\)}}

\put(52,22){\line(0,1){4}} \put(52,22){\circle*{1.4}}
\put(53.5,22){\makebox(0,0)[bl]{\(V^{3}\)}}

\put(50,17){\makebox(0,0)[bl]{\(V^{3}_{1}\)}}
\put(50,13){\makebox(0,0)[bl]{\(V^{3}_{2}\)}}
\put(50,09){\makebox(0,0)[bl]{\(V^{3}_{3}\)}}
\put(50,05){\makebox(0,0)[bl]{\(V^{3}_{4}\)}}

\put(77,22){\line(0,1){4}} \put(77,22){\circle*{1.4}}
\put(78.5,22){\makebox(0,0)[bl]{\(V^{4}\)}}

\put(75,17){\makebox(0,0)[bl]{\(V^{4}_{1}\)}}
\put(75,13){\makebox(0,0)[bl]{\(V^{4}_{2}\)}}

\end{picture}
\end{center}

\newpage
\begin{center}
 {\bf Table 9.} Change operations for \(S^{5G^{adv^{2}}} \Rightarrow  S^{5G^{adv^{2}}}\) (stage 2)   \\
\begin{tabular}{|c|l|c|l|c|c |c| }
\hline
 No. &System component&Alternative&Change/improvement&Binary&Profit&Cost  \\
 \(\iota\)& &under change&item/operation&variable&\(c_{\iota,j}\)&\(b_{\iota,j}\)\\
\hline

 1.& \(B^{32}\) (multiplexing/
  &\(B^{32}_{7}\)&\(V^{1}_{1}:\) None &\(y_{1,1}\)&\(0.0\)&\(0.0\)\\

  &access technology)&\(B^{32}_{7}\)&\(V^{1}_{2}:
   B^{32}_{7} \rightarrow \ B^{32}_{8}\)&\(y_{1,2}\)&\(4.5\)&\(4.0\)\\

\hline
 2.& \(B^{41}\) (core network)
  &\(B^{41}_{7}\)&\(V^{2}_{1}:\) None &\(y_{2,1}\)&\(0.0\)&\(0.0\)\\

  &                 &\(B^{41}_{7}\)&\(V^{2}_{2}:
  B^{41}_{7} \rightarrow \ B^{41}_{8}\)&\(y_{2,2}\)&\(6.5\)&\(7.0\)\\

\hline
 3.& \(B^{441}\) (satellite  network)
  &\(B^{441}_{2}\)&\(V^{3}_{1}:\) None &\(y_{3,1}\)&\(0.0\)&\(0.0\)\\

  &                 &\(B^{441}_{2}\)&\(V^{3}_{2}:
  B^{441}_{2} \rightarrow \ B^{441}_{3}\)&\(y_{3,2}\)&\(6.0\)&\(6.5\)\\

  &                 &\(B^{441}_{2}\)&\(V^{3}_{3}:
  B^{441}_{2} \rightarrow \ B^{441}_{4}\)&\(y_{3,3}\)&\(6.5\)&\(7.5\)\\

  &                 &\(B^{441}_{2}\)&\(V^{3}_{4}:
  B^{441}_{2} \rightarrow \ B^{441}_{5}\)&\(y_{3,4}\)&\(11.0\)&\(18.0\)\\

\hline
 4.& \(B^{442}\) (satellite  functions)
  &\(B^{442}_{1}\)&\(V^{4}_{1}:\) None &\(y_{4,1}\)&\(0.0\)&\(0.0\)\\

  &                 &\(B^{442}_{1}\)&\(V^{4}_{2}:
  B^{442}_{1} \rightarrow \ B^{442}_{2}\)&\(y_{4,2}\)&\(12.0\)&\(30.0\)\\ 

\hline
\end{tabular}
\end{center}


 In general, more complex models can be used at each
 improvement (or forecasting) stage
 (e.g., multicriteria models, combinatorial synthesis approach).
 As a result, several solutions can be obtained and
 the structure of the improvement (forecasting) strategy(ies)
 will be
  more complex
 (e.g., tree, parallel-series graph, network).

\section{Conclusion}

 The article describes an attempt
 to examine the evolution of wireless mobile
 communication technologies/systems
 on the basis of the author combinatorial approach.
 The approach is based on a hierarchical modular model of the
 system under study and an analysis of changes between the system
 generations to collect a set of basic change operations
 (change items).
  The application numerical illustrative examples
 are  based on expert judgment.

 Some future research directions can include the following:
 {\it 1.} analysis of system improvement/evolution (or forecasting)
 for complex applied systems in various domains
 (e.g., software engineering, data based management systems,
 computer engineering);
 {\it 2.} special analysis
 (i.e., improvement, evolution, forecasting processes)
  of subsystems, for example:
  (i) multi-antenna technologies (e.g., MIMO, MU-MIMO),
  (ii) multiple HetNets (at the same area),
  (iii) network architecture/topology, and
  (iv) multiplexing/access technologies;
 {\it 3.} analysis of correlations between two fields:
 (a) improvement/evolution/forecasting processes,
 (b) innovation processes and innovation cycles;
 {\it 4.} designing a special support computer-aided tool
 to implement the described combinatorial approach;
 and
 {\it 5.} usage of the considered approach to structural
 system evolution and forecasting in education
 (CS, applied mathematics, engineering, management).

\section{Acknowledgments}

 This research
 was partially supported by Russian Science Foundation
 grant 14-50-00150 ``Digital technologies and their applications''
 (project of Inst. for Information Transmission Problems).


\end{document}